\documentclass{jpp}
\usepackage{graphicx}

\usepackage[utf8]{inputenc}
\usepackage[T1]{fontenc}
\usepackage{amsmath}

\shorttitle{Hybrid Gyrokinetic Hamiltonian Field Theory}
\shortauthor{F. N. deOliveira-Lopes, et al.}

\title{Geometrical Formulation of Hybrid Kinetic and Gyrokinetic Hamiltonian Field Theory for Astrophysical and Laboratory Plasmas}

\author{F. N. deOliveira-Lopes\aff{1}
  \corresp{\email{Nathan.deOliveira@ipp.mpg.de}},
 D. Told \aff{1} 
 K. Pommois \aff{1}
 K. Hagiwara \aff{1}
 A. Mustonen \aff{1}
 \and R. Grauer \aff{2}}

\affiliation{\aff{1}Max Planck Institute for Plasma Physics, Garching bei Muenchen, Germany.
\aff{2}Ruhr-University Bochum, Bochum, Germany.}

\begin{document}

\maketitle

\begin{abstract}
In the present work, a consistent Lagrangian model that encapsulates fully kinetic ions and gyrokinetic electrons for solar wind electromagnetic turbulence is formulated. Using a consistent method, where both electrons and protons are treated with the same mathematical formalism, we derive and implement a model in which high frequency waves and kinetic electrons effects are described in a computationally cost-efficient way. To that aim, higher order Lie-transform perturbation methods applied to Hamiltonian formulation of guiding center motion are used in order to describe the dynamics of particles and fields. Furthermore, the use of a Hamiltonian formulation allow us to introduce an abelian and gauge invariant electromagnetic field theory for the closure of the system. 
\end{abstract}

\section{Introduction}

Plasma turbulence is germane in astrophysics and space physics (\cite{book_TurbulenceWind, book_SpacePhysics, book_PlasmaAstrophysics}). In weakly magnetized and weakly collisional plasma, empirical data suggests that  kinetic effects play a pivotal role in understanding how energy is dissipated at ion and electron scales (\cite{Alexandrova2009, TurbulenceDissipation, Chen2019}). Space plasma physics is idiosyncratically suitable for the understanding of turbulence because it offers a vast range of frequency scales and the possibility of \textit{in situ} measurements with various spacecraft missions (\cite{Alexandrova2009, PSP_7, PSP_2, PSP_3}). The Parker Solar Probe is the most recent of these missions (\cite{PSP_4, PSP_5}). Early data suggests that even at distances as close as 30 to 50 $R_\odot$, the power spectral density indicates that the transport of energy from large scales into thermal heat is already mediated through kinetic turbulent effects (\cite{PSP_1, PSP_6, PSP_8}).

Energy dissipation in kinetic scales is a yet unresolved problem in space physics (\cite{book_TurbulenceWind}). Currently, the standard picture is that energy is injected in the system in the largest scale, and it is then transported to the smallest scales, where it is transformed to thermal energy (\cite{book_BasicAstro, TurbulenceDissipation}). In the process of dissipation, the statistically isotropic inertial range is elegantly described by Magnetohydrodynamics (\cite{MHD_1, MHD_2}). At the smallest scales, kinetic effects take place (\cite{KAW_1}). Contrary to prior assumption, the temperature of the solar wind at 1AU is higher than adiabatic expansion predicts (\cite{Marsch}, \cite{Kohl}), and in this case, kinetic turbulent effects might be playing an important role.

Particle heating happens in plasma physics due to irreversible thermodynamic processes and the property of time reversal symmetry of physical laws. Together with the well formulated Onsager reciprocity theorem (\cite{Callen}), and the natural tendency of isolated systems to always evolve towards maximum entropy, it allows us to better understand how energy is dissipated. In collisional astrophysical plasma turbulence, the mechanism responsible for energy dissipation in the inertial range is viscous dissipation (\cite{Viscous}). The picture changes in the kinetic scales, where frictional conversion of the bulk energy does not completely explain how energy dissipation takes place. In this case, wave particle interaction should be taken into account \citep{Carbone}. The wave particle interaction forms structures in the velocity space distribution function (\cite{Relaxation}), which in turn increase the collisional energy dissipation efficacy (\cite{Relaxation2}). Among the reversible energy transfer channels between waves and particle, the resonant ones take place when the particle gyrating around the magnetic field resonates with the wave's frequency, and it can happen in the format of Landau damping (\cite{pitaevskii2012physical}), cyclotron damping (\cite{CyclotronDamping}), etc. Non resonant effects takes place in the form of, e. g. stochastic heating or magnetic pumping (\cite{Stochastic}, \cite{Pumping}). Whether or not relaxations in velocity space distribution function explains energy dissipation completely remains to be further investigated. Nevertheless, a more meticulous investigation of irreversible processes must be done in the framework of weakly collisional energy dissipation in order to properly understand the values of solar wind temperature measured (\cite{Verscharen}). In moving forward, it is important to keep in mind previous work that aimed at modeling a similar phenomena to the one we would like to reproduce. The GKe/FKi (\cite{Lin_2005}) was proposed in order to bridge the gap between fluid/kinetic hybrid models and fully kinetic models. Unlike the model proposed in the present work, the parameter space of the derivation is not appropriate for simulating turbulence in space and astrophysical plasma. Furthermore, the derivation of the field equations follows a pull-back/push-forward  approach, whereas the present work uses a variational approach, which we expect to yield better conservation properties. 

In order to better investigate solar wind turbulence at high frequency scales, we are going to develop a theoretical framework focused on turbulent kinetic effects. We are going to derive a hybrid system consisting of a fully kinetic and a gyrokinetic particle species. The gyrokinetic theory is an asymptotic limit of kinetic theory for magnetized plasmas. In this approach, a perturbation expansion is done in order to eliminate theta dependency up to a chosen ordering and make sure that the magnetic momentum is conserved when the background magnetic field is stronger than the various  plasma perturbations. The present work is the first in a series that aims at building a electromagnetic nonlinear hybrid fully kinetic and gyrokinetic model.


In the present work, we make use of a mathematical apparatus available in differential geometry. It is important to highlight here that the use of differential geometric tools not only allow us to derive acceptable energy and momentum conservation properties. It also grant us the possibility to ensure the preservation of the whole hierarchy of structures in the system. Beyond the conservation of energy and momentum, one is also capable of preserving the symplectic capacity, differential forms under Hamiltonian evolution, and conservation laws as derived from the Noether theorem. What is more, it allow us to describe a system that encapsulates organically a geometric description of thermodynamics through contact geometry, the even counterpart of symplectic geometry (\cite{Contact}). 

This work is organized as follows: In Section 2 we are going to perform a detailed geometrical derivation of fully kinetic physics, looking on how the symplectic structure emerges in this theoretical framework and how it allow us to work with the Poisson brackets in order to derive the dynamics of the system. We continue then to section 3, where we work with the truncated equations for the fully kinetic species, and proceed with the derivation of the gyrokinetic equations of motion. It is worth to emphasize here that beyond the phenomenology of the physical problem we are trying tackle, our gyrokinetic equations should be somewhat similar to the gyrokinetics found in the literature. The novelty of the present approach is found on how the fully kinetic equations are structurally connected to the reduced ones and the fields, and how this affects the breadth and scope of the model. In section 4 we construct the gyrokinetic Vlasov equation. After that, in section 5, we derive the field equations with the variational principle, and we finish the work in section 6, with a linear electrostatic analysis.

In 1958, F. E. Low published what was to be known as the first lagrangian formulation of the Boltzmann-Vlasov equations (\cite{Low}). In his work, he emphasized that regardless of no new calculations being performed, the approach would produce a new way to attack practical problems. Inspired by that attitude, the present work does not aim to provide new physics. Nevertheless, we do expect that our hybrid approach allow us to model higher frequency scales, providing a more accurate yet cost-beneficial alternative to standard models.

\section{Geometrical Hamiltonian Mechanics}

In this section, we are going to derive the fully kinetic segment of our model. At the same time, we lay down the theoretical framework used in the present work. One expects that the robustness of the mathematical apparatus used here would help us to tackle more complex problems, such as energy dissipation and heating or, as in the case of gyrokinetics, specific coordinate transformations.

From a physical perspective, our goal is to describe the equations of motion for the particles in the system. That means position and velocity, or position and momentum, and all emergent properties. Mathematically, those quantities represent coordinates on a given space. Here it is important to stress the elementary connection and distinction between mathematics and physics. Mathematics seeks to understand relationships among abstract structures, whereas physics utilizes said structural relationships to investigate the property and behavior of matter and forces through space and time. Henceforth, it is important to understand how this two subjects relate to each other and subsequently coalesce, inasmuch as we base our motivation both in a mathematical and physical argument.

 In the classical non-relativistic case, one establishes a correspondence between a physical coordinate description and operations with geometrical objects on a manifold, where a manifold is a locally-euclidean topological space (\cite{Arnold}, \cite{Was}). The physical reality is represented by coordinates on the configuration space, described in our case on a symplectic manifold. A symplectic manifold is a topological space that allows for the existence of a closed nondegenerated symplectic two-form, $\omega$. Symplectic manifolds are special case of a Poisson manifolds, we shall make use of this property in the following steps.  A differential form is a coordinate independent tensorial description of integrands over a differentiable manifold. In our case, the symplectic two-form allow us to generalize the concept of area from multivariable calculus, to a coordinate free symplectic manifold. The non-degenerancy of the symplectic form allows for the existence of its inverse, which is in turn used for the construction of the Poisson bracket. We are going to describe in more detail this relationships and its ramifications in the following sections.
 
 Beyond the idiosyncratic lexicon, it is important to understand the procedure being described here. For that aim, we wish to use the following section as a toy model to better understand the mathematical framework used in the gyrokinetic derivation and the derivation of the field equations. We aim to build our theoretical framework by constructing a correspondence between elements of our physical system with more general geometric structures.  Such connection will allows us to better understand the fundamental properties of the nature of problem in question.

\subsection{Symplectic and Cosymplectic approach}

In order to derive the equations of motions of the fully kinetic particles,  we need to understand how the symplectic two-form relates to our physical system and how it can be used to extract further information about the phenomena we wish to study. An intuitive way to construct our symplectic matrix, is looking at what it represents geometrically. 

Consider a symplectic vector space $V_{i}$ and a linear map $t\in(V_{i},V_{j})$, where $V$ is a vector space and $L^2$ is the space of all bilinear maps from $V \times V$ to R. A map is considered nondegenerate if for $\omega(v_{1},v_{2})=0$ for all $v_2 \in V$, we must also have $v_1=0$. If we describe the basis and dual basis of the given vector space as $(e_1, ..., e_n)$ and $(e^1, ..., e^n)$, respectively, the map $\omega$ is nondegenerate if the matrix of $\omega_{ij} = w(e_i,e_j)$ is nonsingular. The map is symmetric if its transpose is $\omega^t(e_i,e_j) = \omega(e_i,e_j)$ and skew-symmetric if $\omega^t = -\omega$. Consider now a space of skew-symmetric bilinear maps on V, and within it, a non-degenerated 2-form $\omega \in A^{2}(V)$. In its basis, the matrix $\omega(e_{i},e_{j})$ has the form 

\begin{equation}
J=\left(\begin{array}{cc}
0 & I_{n}\\
I_{n} & 0
\end{array}\right).    
\end{equation}

The pair $(\omega,V)$ is called a symplectic vector space, and a linear diffeomorphism $\phi$ that satisfies $\phi*\omega_2=\omega_1$ is called symplectomorphic for two symplectic 2-forms that transform with $\phi \in L(V_{1},V_{2})$. In the present derivation, we define the phase-space of an autonomous system as a symplectic manifold $(M,\omega)$ and the dynamics of the system is described as one-parameter family of symplectomorphisms. The form used to describe the time evolution of the system on the manifold M is closed, i.e. $d\omega=0$, and exact, i.e. $\omega = d\theta$. If we define $\theta = \Gamma$, where $\Gamma=\mathcal{L}dt$ is our phase space Lagrangian one-form, and the tautological one-form provides a bridge between the Lagrangian and the Hamiltonian mechanics. The use of such a formalism also allow us to gauge transform our Lagrangian, and as far as the gauge functions can be described as a exterior derivatives, i.e. $d\sigma$, the dynamics of the system remains unchanged. This feature will become clear on the gyrokinetic transformation reduction. In this case, the general symplectic map $\phi$ from two symplectic manifolds $(M_1,\omega_1)$ and $(M_2,\omega_2)$ is a canonical transformation if $\phi*\omega_2 = \omega_1$. 

We now consider a phase-space $T^*\mathcal{M}$, cotangent to the configuration space $\mathcal{M}$. Selecting local coordinates on $\mathcal{M}$ $(dq_1,...,dq_n)$, the basis of the cotangent space $T_p^*\mathcal{M}$ is given by $(dq^1,...,dq^n)$. Considering the tautological contact 1-form $\theta \in T_{q}^{*}\mathcal{M}$ formatted as $\theta = p_idq^i$, the local coordinates in $T^*\mathcal{M}$ are $(q_1,...q_n,p_1,...,p_n)$. The Poincare two-form, also know as canonical symplectic form is then written by the symplectisation of $\theta$ 

\begin{equation}
    \omega=dp_{i}\wedge dq^{i},
\end{equation}

Considering a noncanonical coordinates change of the form

\begin{equation}
    p_i = p_i(\Theta^i,t),
\end{equation}

\begin{equation}
    q^i = q^i(\Theta^i,t),
\end{equation}

where $\Theta^a = (x,v)$ is the non canonical phase space, with the transformations $x=q$, and $v=p-\frac{e}{c}A$. We can write our non canonical Lagrangian one-form as 

\begin{equation}\label{first_lagrangian}
    \mathcal{L} = \Lambda_a \frac{\partial \Theta^a }{\partial t} - \bar{H},
\end{equation}

where $\Lambda_a = p_i \frac{\partial q^i}{\partial \Theta^a}$, and $\bar{H} = p_i \frac{\partial q^i }{\partial t} - H(\Theta) = \frac{1}{2m} |p - \frac{e}{c}A(q,t)|^2 - e\phi(q,t)$. Our tautological one-form becomes then $\Gamma = \mathcal{L}dt = (\Lambda_a d \Theta^a  - \bar{H}dt) $, and our symplectic form acquires the form 

\begin{equation}\label{symplectic_matrix}
\omega_{ab} = d\Gamma = \epsilon_{ijk}B_{k}dx^{i}\wedge dx^{j}-m\delta_{ij}dx^{i}\wedge dv^{j}+m\delta_{ij}dv^{i}\wedge dx^{j}. 
\end{equation} \\

Consider now the existence of a smooth function $H \in M$, with a vector field defined by $i_{X_{H}}\omega+dH=0$. Since the existence of $X_H$ is guaranteed by the nondegeneracy of $\omega$, a Hamiltonian mechanical system can be defined on $(M,\omega,H)$, and in canonical coordinates we have

\begin{equation}
    X_{H}=\left(\frac{\partial H}{\partial p_{i}}\frac{\partial}{\partial q^{i}}-\frac{\partial H}{\partial q^{i}}\frac{\partial}{\partial p_{i}}\right).
\end{equation}

The one-parameter family solution that described the dynamics of our system is then an integral curve of $X_H$ if the following set of equations is satisfied

\begin{equation}
    \dot{q}^{i}=\frac{\partial H}{\partial p_{i}},
\end{equation}

\begin{equation}
   \dot{p_{i}}=-\frac{\partial H}{\partial q_{i}}.
\end{equation}

For a given integral curve of the Hamiltonian vector field $g(t)$, we have that $H(g(t))$ is constant in time for an autonomous system, and the flow $f_t$ of the vector field satisfies $H \dot f_t = H \forall t$. Using the Liouville theorem, we have that $f_t^*\omega=\omega$, and $d/dt f_t^*\omega=0$. Our flow is then symplectic and preserves the volume element given by the orientation $V_{\omega}=\frac{(-1)^{\frac{n(n-1)}{2}}}{n!}\omega^{n}$. Since symplectomorphisms preserve the Hamiltonian structure, a map $s$ will be symplectic only if it satisfies $s^*(X_H)=X_{H \circ s}$.

We consider a vector field on a symplectic manifolds to be at least locally Hamiltonian if the Lie derivative of the symplectic form with respect to the vector field X is zero, that means 

\begin{equation}
    \mathcal{L}_{X}\omega=0.
\end{equation}

Insofar as the Lie derivative describe infinitesimal changes of any given tensors with respect to any given vector field, the equation above help us to describe a vector field that locally conserves the symplectic structure. For any given functions on our manifold, $f,g\in(M,\mathbb{R})$, the Lie derivative $\mathcal{L}_{X_{g}}g=\mathcal{-L}_{X_{f}}f$ defines the Poisson Bracket of the following form

\begin{equation*}
\{f,g\}_{\omega}=-\omega(X_{f},X_{g}).    
\end{equation*}

The group of all symplectomorphisms of a symplectic vector space forms a group under composition called the symplectic group $Sp(M,\omega)$, and any symplectmorphic map belonging to $(M,\omega)$ must also preserve the Poisson Bracket structure, which is in fact a more fundamental structure belonging to a Poisson manifold and a Poisson algebra. It is important to highlight that the use of a Poisson geometry emphasizes the Lie structure of the system through the use of a Lie Bracket, or a Poisson Bracket in this specific case, and therefore it is more suitable for the description of a system endowed with a Hamiltonian structure.

The pair $(M, \{,\}_\omega)$ form a Poisson manifold, and the bracket can be described with a contravariant anti-symmetric two-tensor $\Pi$ known as cosymplectic, or rank-2 Poisson tensor 

\begin{equation*}
\Pi^{ab} : T^*M \times T^*M \longrightarrow \mathbb{R}.
\end{equation*}

In local coordinates, $\Pi$ is defined by its matrix elements $\{\Theta^a, \Theta^b\} = \Pi^{ab}(\Theta)$. In this case we have

\begin{equation}
    \Pi(f,g)\equiv\{f,g\}=\sum_{ab}\Pi^{ab}(\Theta) \frac{\partial f}{\partial \Theta^{a}}\frac{\partial g}{\partial \Theta^{b}}.
\end{equation}

Considering that the Poisson bracket must satisfy the Jacobi identify, we observe that the equation

\begin{equation*}
    \{\{\Theta^a, \Theta^b\}, \Theta^c\} + \{\{\Theta^c, \Theta^a\}, \Theta^b\} + \{\{\Theta^b, \Theta^c\}, \Theta^a\} = 0,
\end{equation*}

can be written as 

\begin{equation}
\Pi^{da}\frac{\partial\Pi^{bc}}{\partial\Theta^{d}}+\Pi^{db}\frac{\partial\Pi^{ca}}{\partial\Theta^{d}}+\Pi^{dc}\frac{\partial\Pi^{ab}}{\partial\Theta^{d}}=0.
\end{equation}

If we then write $X_H F = \{F,H\}$, in coordinates we have 

\begin{equation}
X_{H}^{a}\frac{\partial F}{\partial\Theta^{a}}=\Pi^{bc}\frac{\partial F}{\partial\Theta^{b}}\frac{\partial H}{\partial\Theta^{c}}\rightarrow X_{H}^{a}=\Pi^{ab}\frac{\partial H}{\partial\Theta^{b}}.
\end{equation}

This relationship show us that $\Pi$ can be written as the negative inverse of the symplectic matrix, in agreement with the expectation of a nondegenerated case. Writing the symplectic matrix of equation \ref{symplectic_matrix} in the matrix form, our Poisson tensor becomes 

\begin{equation*}
\begin{split}\Pi^{\alpha\beta}=\omega_{\alpha\beta}^{-1}=\begin{pmatrix}0 & \frac{1}{m}\delta^{ij}\\
-\frac{1}{m}\delta^{ij} & \frac{1}{m^{2}}\epsilon^{ijk}B_{k}
\end{pmatrix}\end{split},
\end{equation*}

and our Poisson Bracket becomes 

\begin{equation}
\{f,g\}=\sum_{ab}\frac{\partial f}{\partial \Theta^{a}}\Pi^{ab}\frac{\partial g}{\partial \Theta^{b}}=\frac{1}{m}\left(\nabla f\frac{\partial g}{\partial v}-\frac{\partial f}{\partial v}\nabla f\right)+B\left(\frac{\partial f}{\partial v}\times\frac{\partial g}{\partial v}\right),
\end{equation}

Finally, considering a system $(M,\omega,H)$ with a Hamiltoniann vector field $X_H$ and flow $\phi_t$, and considering also an infinite differentiable function $f$ on this manifold, the equations of motion in the Poisson Bracket form are given by

\begin{equation*}
    \frac{d}{dt}(f \circ \phi_{t})_{\omega}=\{f \circ \phi_{t},H\}_{\omega}=\{f,H\}_{w}.
\end{equation*}

Considering now our local non canonical variable $\Theta = (x,v)$, we have

\begin{equation}
\begin{split}\dot{x}=\{x^{},H\}=\delta^{ij}\frac{\partial H}{\partial v^{j}}=\Pi^{x^{i}x^{j}}\frac{\partial H}{\partial x^{j}}+\Pi^{x^{i}v^{j}}\frac{\partial H}{\partial v^{j}}=v^{}\end{split},
\label{eq45}
\end{equation}

and

\begin{equation}
\begin{split}\dot{v} =\{v^{},H\}=\Pi^{v^{i}x^{i}}\frac{\partial H}{\partial x^{j}}+\Pi^{v^{i}v^{j}}\frac{\partial H}{\partial v^{j}}=-\delta^{ij}\frac{\partial H}{\partial x^{j}}+\epsilon^{ijk}\frac{\partial H}{\partial v^{i}}=\frac{\partial\phi}{\partial x^{i}}+\epsilon^{ijk}v_{j}B_{k}\end{split}.
\label{eq46}
\end{equation}

In order to construct the Vlasov equation, we should evaluate the  Liouville’s Theorem  on the context of symplectic geometry. Considering that the phase-space distribution function $F(p,q)$ is constant along the trajectories of the the system, we can describe the time evolution of a volume element of our phase space according to the the following relation

\begin{equation*}
    \frac{dF}{dt}=\frac{\partial F}{\partial t}+\sum_{i=1}^{n}\left(\frac{\partial F}{\partial q^{i}}\dot{q}^{i}+\frac{\partial F}{\partial p_{i}}\dot{p}_{i}\right)=0.
\end{equation*}

Because the phase space distribution function is constant along trajectories in the phase space, the theorem says that the Liouville measure is invariant under Hamiltonian flows. Geometrically, the relationship becomes 

\begin{equation*}
    \frac{dF}{dt}= \{F ,H \}.\label{liouville}
\end{equation*}

The Vlasov equation becomes  

\begin{equation*}
\begin{split}\frac{\partial F^{}}{\partial t}+\left\{ x^{},H^{}\right\} \nabla F^{}+\left\{ v^{},H^{}\right\} \partial_{v^{}}F^{}=0.\end{split}
\end{equation*}

Substituting the values computed in equations \ref{eq45} and \ref{eq46}
we have the distribution function equation for the fully kinetic species

\begin{equation}
\frac{\partial F_{}^{}}{\partial t}+v^{I}\nabla F_{}^{}+ \left(\frac{\partial\phi}{\partial x^{i}}+\epsilon^{ijk}v_{j}B_{k}\right)\partial_{v^{}}F_{}^{}=0.\label{final_fully}
\end{equation}

It is important to emphasize here that, despite the recognized outcome of this calculation, the use of the geometrical approach allow us to work with a framework which is favourable for a coherent gyrokinetic coordination reduction. This is due to the fact that we start the transformation from the tautological one-form defined by the Lagrangian equation \ref{first_lagrangian}, instead of working directly from the final equations \ref{eq45}, \ref{eq46}, and \ref{final_fully}. The use of such an elegant and robust theoretical framework will also prove itself advantageous when we introduce the concept of geometrical Entropy.

\section{Gyrokinetic Equations of Motion}

The full gyrokinetic coordinate reduction used in the present model is derived in detail in Appendix \ref{appendix_b}. In the present section, the Lagrangian resulting from Appendix \ref{appendix_b} is embedded in the sum of the phase spaces of all the variable required to describe the total dynamics of the system. Now we write down our Lagrangian containing terms from the fully kinetics dynamics, which is represented by $\mathcal{L}^p_{fk,i}$, as well as the gyrokinetics, written as $\mathcal{L}^p_{gy}$, and the electromagnetic field, $\mathcal{L}^f $. \\


Our full one-form gyrokinetic phase space Lagrangian is divided into three parts, with the following form 

\begin{eqnarray}\label{full_lagrangian_2}
\mathcal{L}^p_{fk,i} & = & \int d\Omega_1 \Sigma_i \left\{ \left(m_i \textbf{v}_i+\frac{q_i}{c}\textbf{A}(x,t)\right)\boldsymbol{\boldsymbol{\cdot}}\dot{x}-\frac{1}{2}m_i|\textbf{v}_i|^{2}+q_i\phi_{1}(x,t)\right\}, 
\end{eqnarray}
\begin{multline}\label{gyro_lagrangian}
  \mathcal{L}^p_{gy} =  \int d\Omega_2 \left\{ \left(\frac{e}{\varepsilon_\delta c}\textbf{A}_{0}+m\textbf{v}_{\parallel gy}\hat{\textbf{b}}(\textbf{X}_{gy})\right)\boldsymbol{\cdot}\dot{\textbf{X}}_{gy}+\varepsilon_\delta\frac{mc}{e}\mu_{gy}\dot{\theta}_{gy} \right. \\
  \left. - \frac{1}{2}mv_{\parallel gy}^{2}-\mu_{gy}B(\textbf{X}_{gy})-\varepsilon_{\delta}e\langle\psi_{1}\rangle \right. \\
  \left. - \varepsilon_{\delta}^{2}e^{2}\left(\frac{1}{2mc^{2}}\langle|\textbf{A}_{1}|^{2}\rangle-\frac{1}{2B(\textbf{X}_{gy})}\partial_{\mu_{gy}}\langle\psi_{1}^{2}\rangle\right) \right \},
\end{multline}

\begin{multline}
\left. \mathcal{L}^f = \frac{1}{8\pi}\int_{v}d^{3}x_{}\left(\varepsilon_{\delta}^{2}|\nabla\phi_{1}(x_{})|^{2}-|\nabla\times[\textbf{A}_{0}(x_{}) \right. \right. +\varepsilon_{\delta}\textbf{A}_{1}(x_{},t)]|^{2}+\\
\left. \left. \varepsilon_{\delta}\frac{2}{c}\lambda(x_{},t)\nabla\boldsymbol{\cdot} \textbf{A}_{1}(x_{},t)\right)\right\}.
\end{multline}

Our final Lagrangian can be written as the sum of the previous equations, that is 

\begin{equation}
\mathcal{L} = \mathcal{L}^p_{fk,i} + \mathcal{L}^p_{gy} + \mathcal{L}^f.
\end{equation}

 Here $d\Omega_{1}=d^{3}v_{}d^{3}x_{}$, is the fully kinetic phase space, and $d\Omega_{2}=d^{3}\textbf{X}_{gy}\textbf{B}_{\parallel}^{*}dv_{\parallel gy}d\mu_{gy}d\theta_{gy}$ is the gyrokinetic phase space. In the present section, we are going to focus on the gyrokinetic Lagrangian, equation \ref{gyro_lagrangian}. We also write our fields as

\[
 \textbf{A}_{1}(\textbf{X}_{gy}+\rho)=\textbf{A}_{1} \boldsymbol{\cdot} \hat{c}+ \textbf{A}_{1} \boldsymbol{\cdot} \hat{b}, \\
\]
where $ \textbf{A}_{1\perp} = \sqrt{\frac{2 \mu_{gy}B}{mc^2}}\textbf{A}_{1} \boldsymbol{\cdot} \hat{c}(\theta_{gy},X_{gy})$,  $ {A}_{1\parallel} = \frac{v_{gy, \parallel}}{c}\textbf{A}_{1} \boldsymbol{\cdot} \hat{b}$, and the electric potential is described as $\phi_{1}(\textbf{X}_{gy}+\rho).$

The full perturbed electromagnetic potential becomes 

\begin{equation}
{\psi}_{1}(\textbf{X}_{gy}+\rho)=\phi_{1}-\frac{v_{gy,\parallel}}{c}\textbf{A}_{1}\boldsymbol{\cdot} \hat{b}-\sqrt{\frac{2\mu_{gy}B(\textbf{X}_{gy})}{mc^{2}}} \textbf{A}_{1} \boldsymbol{\cdot} \hat{c}(\theta_{gy},X_{gy}).\label{em_potential_eq}
\end{equation}

It is worth to emphasize here the relationship with the derivation of  the gyrokinetic equations of motion and the classical derivation. Making use of the Poisson bracket derived on the appendix A, we can compute the equations of motions used on the construction of our Vlasov equation. 

Considering our Hamiltonian to take the form

\begin{multline}
H_{gy} = \frac{1}{2}mv_{gy, \parallel}^{2}-\mu_{gy}B(\textbf{X}_{gy})-\varepsilon_{\delta}e\langle  \psi_{1}\rangle \\
-\varepsilon_{\delta}^{2}e^{2}\left(\frac{1}{2mc^{2}}\langle|\textbf{A}_{1}|^{2}\rangle-\frac{1}{2B(\textbf{X}_{gy})}\partial_{\mu_{gy}}\langle\psi_{1}^{2}\rangle\right), 
\end{multline}

the final equations, also known as characteristics, look like 

\begin{equation}
\dot{\textbf{X}}_{gy}=\left\{ \textbf{X}_{gy},H_{gy}\right\} =\frac{B^{*}}{mB_{\parallel}^{*}}\frac{\partial H_{gy}}{\partial v_{gy, \parallel}}+\varepsilon_\delta \frac{c\hat{b}}{eB_{\parallel}^{*}} \times \nabla^{*}H_{gy},
\end{equation}
\begin{multline}
\dot{v}_{gy, \parallel} =\frac{\textbf{B}^{*}}{m{B}_{\parallel}^{*}}\boldsymbol{\cdot}\left(\nabla^*v_{gy, \parallel}\frac{\partial H_{gy}}{\partial v_{gy, \parallel}}-\nabla^{*}H_{gy}\right) \\
-\frac{c\hat{{b}}}{e{B}_{\parallel}^{*}}\varepsilon_\delta\left(\nabla^{*}v_{gy, \parallel}\times\nabla^{*}H_{gy}\right) = -\frac{\textbf{B}^{*}}{m{B}_{\parallel}^{*}}\boldsymbol{\cdot}\left(\nabla^{*}H_{gy}\right),
\end{multline}

\begin{equation}
\dot{\mu}_{gy}=\left\{ \mu_{gy},H_{gy}\right\} =-\frac{e}{mc}\frac{1}{\varepsilon_\delta}\partial_{\theta_{gy}}H_{gy}+\frac{\textbf{B}^{*}}{m{B}_{\parallel}^{*}}\nabla\mu_{gy}-\frac{c\hat{\textbf{b}}}{e{B}_{\parallel}^{*}}\varepsilon_\delta\left(\nabla\mu_{gy}\times\nabla^{*}H_{gy}\right)\label{eq:20},
\end{equation}

which give us in turn

\begin{equation}
\dot{\mu}_{gy}=-\frac{e}{mc}\frac{1}{\varepsilon_\delta}\partial_{\theta_{gy}}H_{gy}=0\label{eq:21},
\end{equation}
and lastly

\begin{equation}
\dot{\theta}_{gy}=\frac{e}{mc}\frac{1}{\varepsilon_\delta}\partial_{\mu_{gy}}H_{gy}+\frac{\nabla^{*}\theta_{gy}}{\nabla^{*}X_{gy}}\boldsymbol{\cdot}\frac{\partial\textbf{X}_{gy}}{\partial t}.
\end{equation}
It is important to remember that for the present model, only $\dot{\textbf{X}}_{gy}$ and $\dot{v}_{gy, \parallel} $ are needed to compute the dynamics of the system, up to the chosen ordering. It is also important to remember that since our system is derived to be theta independent, the equation of motion for this variable does not need to be computed.

\section{Conservation Equation}

In order to derive the final gyrokinetic Vlasov equation, we can write down the conservation form of the distribution function on the phase space $dZ_{gy} = dX_{gy} d\mu_{gy} d v_{gy, \parallel} d\theta_{gy}$. The process is similar to the derivation of the conservation equation derived for the fully kinetic species. Taking in consideration the Jacobian of the gyrokinetic coordinate transformation ${\mathcal{J}}(Z_{gy})=B_{\parallel}^{*}(X_{gy})/m$, the gyrokinetic phase space conservation law can be derived as following

\begin{equation*}
\frac{\partial}{\partial Z_{gy}}\boldsymbol{\cdot}\left[{\cal J}(Z_{gy})\left\{ Z_{gy},H_{gy}(Z_{gy},t)\right\}_{gc} \right]=0
\end{equation*}

Considering that ${\cal J}(W)F(W,t)=\int d^{6}W_{0}{\cal J}(Z_{gy,0})F(Z_{gy,0},t_{0})\delta^{6}[Z_{gy}-Z_{gy}(Z_{gy,0},t_{0};t)]$, and that 

\begin{equation*}
\frac{d Z_{gy} }{dt}=\left\{ Z_{gy} ,H_{gy}(Z_{gy},t)\right\}_{gc} ,
\end{equation*}
the Vlasov equations acquires the form

\begin{equation*}
\frac{\partial}{\partial t}\left[{\cal J}(Z_{gy})F(Z_{gy},t)\right]+\frac{\partial}{\partial Z_{gy}}\boldsymbol{\cdot}\left[{\cal J}(Z_{gy})F(Z_{gy},t)\left\{ Z_{gy},H_{gy}(Z_{gy},t)\right\}_{gc} \right]=0.
\end{equation*}
In a more compact way

\begin{equation*}
\left[\frac{\partial}{\partial t}+\left\{ Z_{gy},H_{gy}(Z_{gy},t)\right\}_{gc} \frac{\partial}{\partial Z_{gy}}\right]F(Z_{gy},t)=0.
\end{equation*}

Our final gyrokinetic Vlasov equation becomes

\begin{equation*}
   \frac{\partial F}{\partial t}+\frac{\textbf{B}^{*}}{m{B}_{\parallel}^{*}}\boldsymbol{\cdot}\left(mv_{gy, \parallel}-\frac{e}{c}\left\langle {A}_{1\parallel}\right\rangle \right)\nabla_{gy}F \\
   \end{equation*}
   \begin{equation*}
   +\frac{c\hat{\textbf{b}}}{e{B}^{*}_{\parallel}}\times\left(\mu_{gy}\nabla_{gy}B(\textbf{X}_{gy})+\varepsilon_{\delta}e\nabla\left\langle \phi_{1}\right\rangle -\varepsilon_{\delta}\frac{e}{c}v_{gy, \parallel}\left\langle \nabla {A}_{1\parallel}\right\rangle \right) \boldsymbol{\cdot} \nabla_{gy}F \\
   \end{equation*}
   \begin{equation}
   -\frac{\textbf{B}^{*}}{m{B}_{\parallel}^{*}} \boldsymbol{\cdot} \left(\mu_{gy}\nabla_{gy}B(\textbf{X}_{gy})+\varepsilon_{\delta}e\nabla\left\langle \phi_{1}\right\rangle -\varepsilon_{\delta}\frac{e}{c}v_{gy, \parallel}\left\langle \nabla {A}_{1\parallel}\right\rangle \right)\partial_{v_{gy, \parallel}}F=0,
\end{equation}

where the second term on the first line is associated with the canonical momentum, while the terms on the second line are the grad B drift, $E \times B$ and magnetic flutter. The last term is associated with the total parallel acceleration.

\section{Field Equations}

In order to derive the field equations of our system, we start from the principle that our system lies on a heterogeneous manifold (\cite{Leok}), which is understood to be heterogeneous because it serves as the phase space for our fully kinetic ions and electromagnetic fields, as well as for our coordinate reduced gyrokinetic electrons. The missing piece of our derivation is what connects the fully kinetic and the gyrokinetic dynamics, i. e. the electromagnetic field.  Here, we consider that the field Lagrangian lives in the same manifold as the fully kinetic particles. 

The  approach taken for the derivation of the field equations relies on the use of the variational principle. More specifically, we are going to use the following relationship, known as the functional derivative 

\begin{equation}
    \frac{\delta\mathcal{S}[\chi(\Omega)]}{\delta\chi(\Omega)}\circ\hat{\chi}(\Omega)=0.
\end{equation}

Here,  $\mathcal{S}$ stands for the action of the system, $\chi(\Omega)$ is the field with which we are varying our system, and $\hat{\chi}(\Omega)$ is a testing field. For a detailed derivation of the functional derivative in the context of heterogeneous manifolds, we refer the reader to (\cite{Nathan}). Using the same notation described in Appendix \ref{appendix_b}, and considering that the electromagnetic field does not lie in the same manifold of our gyrokinetic system, i.e. $(\Omega=(x,v)$, and $\Omega_{gy}=(\mathbf{X}_{gy},\mathbf{v}_{\parallel gy},\mu_{gy},\theta))$, in order to perform the variation on the same coordinate system, we must make a transformation such as

\begin{equation*}
\frac{\delta\phi_{1}(X_{gy}+\rho)}{\delta\phi_{1}(\textbf{x})}\circ\hat{\chi}_{1}(\textbf{x})=\frac{\delta}{\delta\phi_{1}(\textbf{x})}\left(\phi_{1}(X_{gy}+\rho)\right)\circ\hat{\chi}_{}(\textbf{x})
\end{equation*}

\begin{equation*}
=\frac{\delta}{\delta\phi_{1}(\textbf{x})}\left[\int d^{3}x\phi_{1}(\textbf{x})\delta^{3}(X_{gy}+\rho-x)\right]\circ\hat{\chi}_{}(\textbf{x})
\end{equation*}

\begin{equation*}
=\frac{d}{d\nu}\left[\int \int d^{3}x\left(\phi_{1}(\textbf{x})+\nu\hat{\chi} (\textbf{x})\right)\delta^{3}(X_{gy}+\rho-x)\right]\left|_{\nu=0}\right.
\end{equation*}

\begin{equation}
=\frac{d}{d\nu}\left[\phi_{1}(X_{gy}+\rho)+\nu\hat{\chi}_{}(X_{gy}+\rho)\right]\left|_{\nu=0}\right.=\hat{\chi}_{}(X_{gy}+\rho). 
\end{equation} \\

Our heterogeneous action is then divided in three parts, the gyrokinetic, fully kinetic and field action, that is $\mathcal{S}_{gy}^{p}$, $\mathcal{S}_{fk}^{p}$, and $\mathcal{S}^{f}$, respectively. In order to derive the complete field equations, we need to perform the variation of all actions with respect to the different fields.

Considering the Lagrangian described in the beginning of section 3, that is

\begin{equation}
\mathcal{L} =  \mathcal{L}^p_{gy} + \mathcal{L}^p_{fk,i} + \mathcal{L}^f,
\end{equation}

the total action of our system is given in a compact form as the time integral of the Lagrangian density, which for various ion species give us

\begin{equation}
\mathcal{S} = \mathcal{S}_{gy}^{p} + \sum_i \mathcal{S}_{fk,i}^{p} + \mathcal{S}^{f}
\end{equation}
where

\begin{multline}
\mathcal{S}_{gy}^{p}=\int\int F_e(\textbf{X}_{gy},v_{gy, \parallel},\mu_{gy},t)\left\{ \left(\frac{e}{ \varepsilon_\delta c}\textbf{A}_{0}+mv_{gy, \parallel}\hat{b}(\textbf{X}_{gy})\right)\boldsymbol{\cdot}\dot{X}_{gy}   \right. \\ \left. +\varepsilon_\delta \frac{mc}{e}\mu_{gy}\dot{\theta}_{gy}- \frac{1}{2}mv_{gy, \parallel}^{2}-\mu_{gy}B(\textbf{X}_{gy})- \varepsilon_\delta e\langle\psi_{1}\rangle \right. \\ \left. - \varepsilon_\delta^2 e^{2}\left(\frac{1}{2mc^{2}}\langle|\textbf{A}_{1}|^{2}\rangle-\frac{1}{2B(\textbf{X}_{gy})}\partial_{\mu_{gy}}\langle \tilde{\psi}_{1}^{2}\rangle\right)\right\} dtd\Omega_{gy},
\end{multline}
\begin{multline}
\mathcal{S}_{fk, i}^{p}=\int\int F_i(x,v)\left\{ \left(m_{0}\textbf{v}+\frac{q_i}{c}[\textbf{A}_{0}(\textbf{x})+\varepsilon_{\delta}\textbf{A}_{1}(\textbf{x})]\right)\boldsymbol{\cdot}\dot{x}\right. \\
\left. -\frac{1}{2}m_{0}|\textbf{v}|^{2}+ q_i\phi(\textbf{x})\right\} dtd\Omega,
\end{multline}
and 
\begin{multline}
\mathcal{S}^{f}=\int\left\{ \frac{1}{8\pi}\int_{v}d^{3}x\left(\varepsilon_{\delta}^{2}|\nabla\phi_{1}(\textbf{x})|^{2}-|\nabla\times[\textbf{A}_{0}(\textbf{x})+\varepsilon_{\delta}\textbf{A}_{1}(\textbf{x})]|^{2}\right. \right. \\
\left. \left. +\varepsilon_{\delta}\frac{2}{c}\lambda(\textbf{x})\nabla\boldsymbol{\cdot} \textbf{A}_{1}(\textbf{x})\right)\right\} dt.
\end{multline}
An important requirement of perturbation theory is the proper understanding of the different orderings imposed on the system and how they affect the final set of equations one aims to implement numerically. In the next sections, we perform the variation of the full action with respect to the different field potentials in order to derive the field equations of our hybrid system.

\subsection{Poisson Equation}

We start with the derivation of the Poisson equation. Applying the tools discussed on the previous section, we perform the variation of the actions with respect to the the electric potential. 

In order to derive the gyrokinetic Poisson equation, one needs to perform the variation with respect to the perturbed electric potential, i.e.  $\delta\mathcal{S}/\delta\phi_{1}$. In this case, all terms $\phi_{1}$-independent vanish, and we are left with the contributions from the electric part of the electromagnetic potential $\psi_{1}$, and the electric contribution from the fully and gyrokinetic action. We are going to first perform the variation with respect to the perturbed electric potential on the gyrokinetic action. 

We start with the derivation of the standard Poisson equation in vacuum using the functional derivative approach on the field action.  

\[
\frac{\delta \mathcal{S}^{f}}{\delta\phi_{1}(\textbf{x})}\circ\hat{\chi}(\textbf{x}) =
\]

\begin{multline*}
\frac{d}{d \nu} \left.  \left[\int \left\{ \frac{1}{8\pi}\int_{v}d^{3}x  \left( \varepsilon_{\delta}^{2}|\nabla\phi_{1}(\textbf{x})|^{2}   \right. \right. \right. \right. \\
-| \left. \left. \left. \left.  \nabla\times[\textbf{A}_{0}(\textbf{x})+\varepsilon_{\delta}\textbf{A}_{1}(\textbf{x})]|^{2} + \varepsilon_{\delta}\frac{2}{c}\lambda(\textbf{x})\nabla\boldsymbol{\cdot} \textbf{A}_{1}(\textbf{x})  \right) \right\} dt \right] \right\vert_{\nu=0} \circ \hat{\chi}(\textbf{x})
\end{multline*}

\begin{equation*}
= \frac{d}{d \nu}\left.\left[\int\left\{ \frac{1}{8\pi}\int_{v}d^{3}x\varepsilon_{\delta}^{2}|\nabla \left (\phi_{1}(\textbf{x})+\nu\hat{\chi}(\textbf{x})\right)|^{2}\right\} dt\right]  \right\vert_{\nu=0}
\end{equation*}

\begin{equation}
= \int\left\{ \frac{\varepsilon_{\delta}^{2}}{4\pi}\int_{v}d^{3}x\nabla \phi_{1}(\textbf{x}) \boldsymbol{\cdot} \nabla \hat{\chi} (\textbf{x}) \right\} dt.\label{eq12}
\end{equation}

Integrating equation \ref{eq12} by parts, and requiring that the boundary term on the integral is zero at infinity, and also taking in consideration the standard gyrokinetic ordering, i.e. $k_\parallel \ll k_\perp$, we are left with \\

\begin{equation}
\frac{\delta \mathcal{S}^{f}}{\delta\phi_{1}(\textbf{x})}\circ\hat{\chi}(\textbf{x}) = - \int\left\{ \frac{\varepsilon_{\delta}^{2}}{4\pi}\int_{v}d^{3}x\nabla_\perp ^2 \phi_{1}(\textbf{x}) \hat{\chi} (\textbf{x}) \right\} dt. 
\end{equation}

In order to maintain consistency, the fully kinetic species is also derived using the functional derivative approach. For the action in consideration, we have

\[
\frac{\delta \mathcal{S}_{fk, i}^{p}}{\delta\phi_{1}(\textbf{x})}\circ\hat{\chi}(\textbf{x}) = \left. q_i \frac{d}{d \nu} \left\{ \int \int F_i(x,v) \left (\phi_{1}(\textbf{x})+\nu\hat{\chi}(\textbf{x})\right) \right\} dt d\Omega \right\vert_{\nu=0}
\]

\begin{equation}
= q_i  \int \int F_i(x,v) \hat{\chi}(\textbf{x}) dt d\Omega.
\end{equation}

Now we perform the full variation of the gyrokinetic action $\mathcal{S}_{gy}^{p}$ with respect to $\phi_{1}(\textbf{x})$

\begin{equation}
\frac{\delta}{\delta\phi_{1}(\textbf{x})}\mathcal{S}_{gy}^{p}\circ\hat{\chi}(\textbf{x})= \int \frac{\delta}{\delta\phi_{1}(\textbf{x})} d\Omega_{gy} dt\mathcal{H}F_e.
\end{equation}

For convenience,  we will express $F_e(X_{gy},v_{\parallel gy},\mu_{gy},\theta_{gy})$ as $F_e$, $\mathcal{H}(X_{gy},v_{\parallel gy},\mu_{gy},\theta_{gy})$ as $\mathcal{H}$, and $\psi_{1}(X_{gy}+\rho)$ as $\psi_1$, unless otherwise explicitly stated. The Hamiltonian part of the action is  

\begin{multline}
\mathcal{H} = \frac{1}{2}mv_{gy, \parallel}^{2}-\mu_{gy}B(\textbf{X}_{gy})- \\
\varepsilon_{\delta}e\langle\psi_{1}\rangle -\varepsilon_{\delta}^{2}e^{2}\left(\frac{1}{2mc^{2}}\langle|\textbf{A}_{1}|^{2}\rangle-\frac{1}{2B(\textbf{X}_{gy})}\partial_{\mu_{gy}}\langle \tilde{\psi}_{1}^{2}\rangle\right),
\end{multline}

and the total variation with respect to $\phi_1(\textbf{x})$ is then

\begin{equation*}
\frac{\delta}{\delta\phi_{1}(\textbf{x})} \int d\Omega_{gy} dt \left[(-\varepsilon_{\delta}e\left\langle \psi_{1}\right\rangle ) -\frac{\varepsilon_{\delta}^{2}e^{2}}{2B(X_{gy})}\partial_{\mu_{gy}} \left\langle \tilde{\psi}_{1}^{2} \right\rangle \right]F_e \circ\hat{\chi}(\textbf{x})\label{eq:43}.
\end{equation*}

We are going to divide the equation \ref{eq:43} in two parts. First we are going to perform the variation on the first integral, referent to $\mathcal{H}(\mathcal{O}(\varepsilon_{\delta}))$, by making use of the tools described earlier. In this case we have

\begin{equation}
- \frac{\delta}{\delta\phi_{1}(\textbf{x})}\int d\Omega_{gy} dt \varepsilon_{\delta} e \left\langle \psi_{1}\right\rangle F_e =-\frac{e\varepsilon_{\delta}}{2\pi} \frac{\delta}{\delta\phi_{1}(\textbf{x})} \int \int d\theta d^{3}x dW  dt \psi_{1}(\textbf{x}) F_e(x-\rho),
\end{equation}

where we considered that $d\Omega_{gy}=d^{3}X_{gy}dW$, and also that $\left\langle \boldsymbol{\cdot} \right\rangle = \frac{1}{2\pi} \int_0^{2\pi} d\theta$. Applying the functional derivative we have then

\begin{equation*}
\left. -\frac{e\varepsilon_{\delta}}{2\pi}\frac{d}{d\nu}\left [ \int d^{3}xdWd\theta dt(\psi_{1}(\textbf{x})+\nu\hat{\chi}(\textbf{x})) F_e(x-\rho)  \right ] \right\vert_{\nu=0}
\end{equation*}

\begin{equation}
=- e\varepsilon_{\delta} \int d^{3}x dt d v_{gy, \parallel}d\mu_{gy} \left\langle F_e(x-\rho)\right\rangle \hat{\chi}(\textbf{x}),
\end{equation}

where $\left\langle F_e(x-\rho)\right\rangle$ corresponds to the gyroaveraged electron distribution function, which in our case, after the integration, becomes the electron gyrocenter charge density. 

We now proceed with the second part of equation \ref{eq:43}, corresponding to the second order correction on the perturbation of the Hamiltonian. The Hamiltonian  is reduced to

\begin{equation}
\mathcal{H}(\mathcal{O}(\varepsilon^2_{\delta})) = -\frac{e^{2}}{2B(X_{gy})}\partial_{\mu_{gy}}\left\langle \tilde{\psi}_{1}^{2}\right\rangle.
\end{equation}

In order to perform the variation on $\left\langle \tilde{\psi}_{1}^{2}\right\rangle$, we first carry out a low wave number expansion. We choose to do so, due to the fact that we want to focus on ion spatial scales in the present model.  This approximation give us

\begin{equation}
\partial_{\mu_{gy}}\left\langle \tilde{\psi}_{1}^{2}\right\rangle \simeq-\frac{m_ec^2}{e^2B(X_{gy})}\left|\nabla_{\perp}\psi_1(X_{gy})\right|^{2}.\label{eq.wave_expansion}
\end{equation}

The details of the low wave number expansion can be found in Appendix \ref{appendix_b}. We are left with

\begin{equation}
 \mathcal{H}(\mathcal{O}(\varepsilon^2_{\delta}))\simeq-\frac{ m_ec^2}{2B^{2}(X_{gy})}\left|\nabla_{\perp}\psi_{1}(X_{gy})\right|^{2}. 
\end{equation}

We can now proceed with the second part of the variation of the gyrokinetic action associated with $\mathcal{H}(\varepsilon_{\delta}^2)$, i. e. 

\begin{equation*}
\frac{\delta}{\delta\phi_{1}(\textbf{x})} \left [ \int \int  F_e(X_{gy},v_{gy, \parallel},\mu_{gy})\left\{ \left( \varepsilon_{\delta}^{2}e^{2}  \frac{1}{2B(\textbf{X}_{gy})}\partial_{\mu_{gy}}\langle \tilde{\psi}_{1}^{2}(X_{gy} + \rho)\rangle \right) \right\} dt d\Omega_{gy} \right ].
\end{equation*}

Using the low wave number expansion \ref{eq.wave_expansion}, we have

\begin{equation*}
\frac{\delta}{\delta\phi_{1}(\textbf{x})} \left [ \int \int  F_e(X_{gy})\left\{ \left( e^2 \varepsilon_{\delta}^{2}  \frac{m_ec^2}{2B^{2}(X_{gy})}\left|\nabla_{\perp}\psi_{1}(X_{gy})\right|^{2} \right) \right\} dt d\Omega_{gy} \right ] 
\end{equation*}

\begin{equation}
= \varepsilon_{\delta}^{2}e^{2}  \int \int d^3x F_e(\textbf{x}) \left( \frac{m_ec^2}{2B^{2}(\textbf{x})} \nabla_{\perp}\hat{\chi}(\textbf{x}) \boldsymbol{\cdot} \nabla_{\perp}\psi_{1}(\textbf{x}) \right)  dt d\Omega_{gy} . \label{d}
\end{equation}

This step of the derivation is delicate and merits a more detailed explanation. Considering that the electromagnetic potential contain terms depending on $v_\perp$ and $v_\parallel$, it is natural to assume that the integration in velocity space yields special prefactors. For the terms associated with the magnetic potential, a current term arises, and a gyrokinetic polarization density factor appears for the term associated with the electric potential. This is explicitly shown  in equation \ref{final_ampere_eq}. Equation \ref{d} becomes

\begin{equation}
- \varepsilon_{\delta}^{2}  \int \int d^3x \nabla_\perp \boldsymbol{\cdot} F_e(\textbf{x}) \left( \frac{m_ec^2}{2B^{2}(\textbf{x})} \nabla_{\perp}\Psi_{1}(\textbf{x}) \hat{\chi}(\textbf{x})   \right)  dt d\Omega_{gy} ,
\end{equation}

and integrating $F_e(\textbf{x})$ with respect to the velocity, where $\Psi_1 = \int  \psi_1 dv $, we have then

\begin{equation}
- \varepsilon_{\delta}^{2}  \int \int d^3x \nabla_\perp \boldsymbol{\cdot}  \left( \frac{n_e m_ec^2}{2B^{2}(\textbf{x})} \nabla_{\perp}\Psi_{1}(\textbf{x}) \hat{\chi}(\textbf{x})   \right)  dt \label{eq.n.32}.
\end{equation}
If we consider that 

\begin{equation*}
 \frac{n_e m_e c^{2}}{2B^{2}} = \frac{4 \pi n_e m^2 c^2 }{8 \pi m e^2 B^2} = \frac{1}{8\pi}\frac{\omega_{pe}^2}{\Omega_{ce}^2},
\end{equation*}

equation \ref{eq.n.32} becomes then 

\begin{equation*}
 - \frac{1}{8\pi} \varepsilon_{\delta}^{2}  \int \int d^3x \nabla_\perp \boldsymbol{\cdot}  \left( \frac{\omega_{pe}^2}{\Omega_{ce}^2} \nabla_{\perp}\Psi_{1}(\textbf{x}) \hat{\chi}(\textbf{x})   \right)  dt.
\end{equation*}

Considering the frequencies to be time and space-independent, we have 

\begin{equation}
 - \frac{1}{8\pi} \varepsilon_{\delta}^{2} \frac{\omega_{pe}^2}{\Omega_{ce}^2} \int \int d^3x   \nabla_{\perp}^2 \Psi_{1}(\textbf{x}) \hat{\chi}(\textbf{x})    dt.
\end{equation}

Considering a constant density and background magnetic field, the full variation of the action with respect to $\phi_{1}(\textbf{x})$ is

\[
\frac{\delta}{\delta\phi_{1}(\textbf{x})}\mathcal{S}^{f}\circ\hat{\chi}+\frac{\delta}{\delta\phi_{1}(\textbf{x})}\mathcal{S}_{gk}^{pt}\circ\hat{\chi}+\frac{\delta}{\delta\phi_{1}(\textbf{x})}\mathcal{S}_{fk}^{pt}\circ\hat{\chi}=
\]\begin{multline*}
- \int \frac{\varepsilon_{\delta}^{2}}{4\pi}\int_{v}d^{3}x\nabla ^2 \phi_{1}(\textbf{x}) \hat{\chi} (\textbf{x})  dt  -e\varepsilon_{\delta} \int d^{3}xdW dt \left\langle F_e(x-\rho)\right\rangle \hat{\chi}(\textbf{x}) \\
-\frac{1}{8\pi} \varepsilon_{\delta}^{2} \frac{\omega_{pe}^2}{\Omega_{ce}^2} \int \int d^3x   \nabla_{\perp}^2 \Psi_{1}(\textbf{x}) \hat{\chi}(\textbf{x})dt + q_i  \int \int F_i(x,v) \hat{\chi}(\textbf{x}) dt d\Omega = 0.
\end{multline*}

If we consider that $\int dWF_e(\textbf{x})= n_e(\textbf{x})$, $\frac{mc^{2}}{2B^{2}(X_{gy})} n_e(\textbf{x})=\frac{1}{8\pi}\frac{\rho_{th}^{2}}{\lambda_{D}^{2}}=\frac{\omega_{pe}^2}{\Omega_{ce}^2}$, and assuming that

\begin{equation*}
\left\langle F_e(x-\rho)\right\rangle =F_e(\textbf{x})+\mathcal{O}(\varepsilon_{\delta}^{}), 
\end{equation*}

our Poisson equation in the strong form can be written as

\begin{equation*}
\frac{1}{4\pi}\nabla ^2 \phi_{1}(\textbf{x}) +  \frac{\rho_{th}^{2}}{\lambda_{D}^{2}} \nabla_{\perp}^2 \Psi_{1}(\textbf{x}) = \sum_i q_i n_i(\textbf{x}) - e n_e(\textbf{x}).
\end{equation*}

In order to visualize the relationship between the Poisson and parallel Ampere equation,  we describe the electromagnetic potential explicitly. In terms of the full electromagnetic potential, we have then

\begin{equation}
\frac{1}{4\pi}\nabla_\perp ^2 \phi_{1}(\textbf{x}) +  \frac{\rho_{th}^{2}}{\lambda_{D}^{2}} \left (  \nabla_{\perp}^2 \phi_{1}(\textbf{x}) + u_{e \parallel} \nabla_{\perp}^2 A_{1 \parallel}(\textbf{x}) \right ) = \sum_i q_i n_i(\textbf{x}) - e n_e(\textbf{x}),\label{final_ampere_eq}
\end{equation}
which can be rewritten as 

\begin{equation}
\frac{1}{4\pi}\nabla_{\perp}^{2}\phi_{1}\left(1+4\pi\frac{\rho_{th}^{2}}{\lambda_{D}^{2}}\right) + u_{e \parallel} \frac{\rho_{th}^{2}}{\lambda_{D}^{2}} \nabla_{\perp}^2 A_{1 \parallel}(\textbf{x})  = \sum_i q_i n_i(\textbf{x}) - e n_e(\textbf{x}).
\end{equation}

\subsection{Parallel Ampere Equation}

We proceed now with the derivation of the parallel Ampere equation from the action integral. In this section, only terms depending on the magnetic potential are computed. Using the variational derivative approach, we first derive the homogeneous solution of the electromagnetic action

 \[
\frac{\delta}{\delta \textbf{A}_{1}(\textbf{x})}{\cal L}_{f}\circ\hat{\chi}_{1}(\textbf{x})= -\frac{\delta}{\delta \textbf{A}_{1}}\int\left \{ \frac{1}{8\pi}\left|\nabla\times[\textbf{A}_{0}(\textbf{x})+\varepsilon_{\delta}\textbf{A}_{1}(\textbf{x})]\right|^{2}\right. \\
\]

\begin{equation}
+\left. \varepsilon_{\delta}\frac{2}{c}\lambda(x,t)\nabla\boldsymbol{\cdot}\textbf{A}_{1}(\textbf{x})\right\} \circ\hat{\chi}_{1}(\textbf{x})d\Omega.
\end{equation}

We first derive the Coulomb gauge condition, by performing a variation with respect to $\lambda(\textbf{x})$, where

\begin{equation*}
\frac{\delta}{\delta\lambda(\textbf{x})}{\cal L}_f\circ\hat{\chi}_{1}(\textbf{x})=\frac{\partial}{\partial\nu}\left[\intop\frac{1}{8\pi}\int_{\nu}d^{3}x\left(\varepsilon_{\delta}\frac{2}{c}\left(\lambda(\textbf{x})+\nu\hat{\chi}_{1}(\textbf{x})\right)\nabla\boldsymbol{\cdot}\textbf{A}_{1}(\textbf{x})\right)dt\right]|_{\nu=0}
\end{equation*}

\begin{equation*}
\hat{\chi}_{1}(\textbf{x})\nabla\boldsymbol{\cdot}\textbf{A}_{1}(\textbf{x})=0\therefore\nabla\boldsymbol{\cdot}\textbf{A}_{1}(\textbf{x})=0.
\end{equation*}

We now proceed with the variation of the action with respect to ${A}_{1 \parallel}(\textbf{x}) =  \hat{b} \boldsymbol{\cdot} \textbf{A}_1(\textbf{x})$. Notice that the generality of the derivation of the Coulomb gauge condition does not affect the results of the following, parallel derivation. We start with the field lagrangian

\begin{equation}
\frac{\delta}{\delta {A}_{ 1 \parallel}(\textbf{x})}S^{f}\circ\hat{\chi}_{1}(\textbf{x})=\frac{\partial}{\partial\nu}\left[\intop\frac{1}{8\pi}\int_{\nu}d^{3}x\left|\varepsilon_{\delta}\nabla\times\left({A}_{1 \parallel}(\textbf{x})+\nu\hat{\chi}_{1}(\textbf{x})\right)\right|{}^{2}\right],
\end{equation}

where we evolve part of the right hand side as,

\begin{equation*}
 = \varepsilon_{\delta}^{2}\frac{\partial}{\partial\nu} \left. \int_\nu \left [ \left|\nabla\times\left({A}_{1 \parallel}(\textbf{x})+\nu\hat{\chi}_{1}(\textbf{x})\right)\right|^{2} \right ] \right |_{\nu=0}
\end{equation*}

\begin{equation}
=2\varepsilon_{\delta}^{2}\nabla\times{A}_{1 \parallel}(\textbf{x})\boldsymbol{\cdot} \nabla\times\hat{\chi}_{1}(\textbf{x}).
\end{equation}

Without loss of generality, we can consider that $\nabla\times \textbf{A}_{1}=\textbf{B}_{1}$, and also that

\begin{equation*}
\int\int\int \textbf{B}_{1}\boldsymbol{\cdot}(\nabla\times\hat{\chi}_{1})dV=\int\int\int\hat{\chi}_{1}\boldsymbol{\cdot}(\nabla\times \textbf{B}_{1})dV-\oint(\textbf{B}_{1}\times\hat{\chi}_{1})\boldsymbol{\cdot} dS.
\end{equation*}
Considering that the second term on the right hand side vanishes on the boundary, we have then

\begin{equation*}
\frac{\delta}{\delta A_{1\parallel}(\textbf{x})}S^{f}\circ\hat{\chi}_{1}(\textbf{x})=-2\varepsilon_{\delta}^{2}\intop\frac{1}{8\pi}\int_{\nu}d^{3}x(\nabla\times \textbf{B}_{1}(\textbf{x}))\boldsymbol{\cdot} \hat{\chi}_{1}(\textbf{x})dt
\end{equation*}

\begin{equation*}
=-\varepsilon_{\delta}^{2}\intop\frac{1}{4\pi}\int_{\nu}d^{3}x(\nabla\times \textbf{B}_{1}(\textbf{x}))\boldsymbol{\cdot} \hat{\chi}_{1}(\textbf{x})dt=0.
\end{equation*}

In terms of the magnetic potential, taking in consideration the Coulomb gauge condition derived above, and assuming $\nabla\times(\nabla\times\textbf{A}_{1})=\nabla(\nabla\boldsymbol{\cdot}\textbf{A}_{1})-\nabla^{2}\textbf{A}_{1}$, we are left with
\begin{equation}
\varepsilon_{\delta}^{2}\intop\frac{1}{4\pi}\int_{\nu}d^{3}x\nabla_\perp^{2}{A}_{1 \parallel}\hat{\chi}_{1}(\textbf{x})dt=0.
\end{equation}

Here we considered that all fields vanishes on the boundary. The equation above is the Ampere law derived from the nonrelativistic Maxwell action, which means we still need to work out the gyrokinetic terms.  The contribution of the gyrokinetic electrons is then given by

\[
\frac{\delta}{\delta {A}_{1\parallel}(\textbf{x})}\mathcal{S}_{gk}^{p}\circ\hat{\chi}=-\frac{\delta}{\delta {A}_{1 \parallel}(\textbf{x})}\left\{ \int F_{e}\left[\varepsilon_{\delta}e\left\langle \psi_{1}\right\rangle +\varepsilon_{\delta}^{2}e^{2}\left(\frac{1}{2mc^{2}}\left\langle |{A}_{1 \parallel}|^{2}\right\rangle -\right.\right.\right.
\]
\begin{equation}
\left.\left.\left.\frac{1}{2B(X_{gy})}\partial_{\mu}\left\langle \tilde{\Psi_{1}^{2}}\right\rangle \right)\right]dtd\Omega_{gy}\right\} \circ\hat{\chi} \label{eq11}.
\end{equation}

Similarly to the Poisson derivation, and considering $\left\langle \psi_{1}\right\rangle =\bigl\langle\phi_{1}(X_{gy}+\rho)-\frac{1}{c}v_{\parallel gy}(X_{gy}){A}_{1 \parallel}(X_{gy}+\rho)\bigr\rangle$, the first term on the right hand side becomes 

\begin{equation}
\varepsilon_{\delta}e \frac{1}{c}\int d^{3}xdtdv_{gy, \parallel}d\mu_{gy}v_{gy, \parallel}\left\langle F_{e}(\mathbf{x}-\mathbf{\mathbf{\rho}})\right\rangle \hat{\chi}(\textbf{x}).
\end{equation}

The second term on the right hand side of equation \ref{eq11} is simplified using the chain rule inside the gyroaveraging. The second term on the right hand side of equation \ref{eq11}  becomes 

\begin{equation*}
\frac{\varepsilon_{\delta}^{2}e^{2}}{2mc^{2}}\frac{\delta}{\delta A_{1\parallel}(\textbf{x})}\int dtd\Omega_{gy}F_{e}(X_{gy})\left\langle | {A}_{1\parallel}(X_{gy}+\rho)|^{2}\right\rangle \circ\hat{\chi} 
\end{equation*}

\begin{equation}
=\frac{\varepsilon_{\delta}^{2}e^{2}}{mc^{2}}\int d^{3}xdtdv_{gy, \parallel}d\mu_{gy}\bigl\langle F_{e}(x-\rho)\bigr\rangle {A}_{1\parallel}(\textbf{x})\hat{\chi}(\textbf{x}),
\end{equation}

where the last term on the last line contributes to the gyrokinetic current. We now proceed to the last term on equation \ref{eq11}, which is evolved using the same approximation used on the derivation of the Poisson equation, cf. Appendix \ref{appendix_b}, that is

\begin{multline*}
    \varepsilon_{\delta}^{2}e^{2}\frac{\delta}{\delta A_{1\parallel}(\textbf{x})}\left[\int\int d^{3}x\delta(x-X_{gy})F_{e} \right. \\ 
    \left. \left\{ \left(\frac{1}{2B(X_{gy})}\frac{1}{\Omega_{e}} \frac{c}{e}\left|\nabla_{\perp}\psi_{1}(X_{gy},v_{gy, \parallel})\right|^{2}\right)\right\} dtd\Omega_{gy}\right]
\end{multline*}

\begin{equation}
=-\frac{\varepsilon_{\delta}^{2}ec}{\Omega_{e}}\int\int d^{3}xF_{e}(x,v_{gy, \parallel})\frac{1}{B(\textbf{x})}\nabla_{\perp}\psi_{1}(x,v_{gy, \parallel}) \boldsymbol{\cdot} \frac{v_{gy, \parallel}}{c}  \nabla_{\perp}\hat{\chi}(\textbf{x})dtd\Omega_{gy}.
\end{equation}

Performing an integration by parts, we are left with 

\begin{equation*}
\frac{\varepsilon_{\delta}^{2}ec}{\Omega_{e}}\int\int d^{3}x\nabla_{\perp}\boldsymbol{\cdot} \left( F_{e}(x,v_{gy, \parallel})\frac{1}{B(\textbf{x})}\right) \nabla_{\perp}\frac{v_{gy, \parallel}}{c}\psi_{1}(x,v_{gy, \parallel})\hat{\chi}(\textbf{x})dtd\Omega_{gy}.
\end{equation*}

This equation is further divided in two parts, one depending on $\phi_1$ and the other one depending on $A_{1\parallel}$. The $\phi_1$ dependent equation becomes

\begin{equation*}
\frac{\varepsilon_{\delta}^{2}ec}{\Omega_{e}}\int\int d^{3}x\nabla_\perp\boldsymbol{\cdot} \frac{v_{gy, \parallel}}{c} F_{e}(x, v_{gy, \parallel})\frac{1}{B(\textbf{x})}\nabla_{\perp} \phi_{1}\hat{\chi}(\textbf{x})dtd\Omega_{gy}
\end{equation*}

\begin{equation}
= \varepsilon_{\delta}^{2}\int\int d^{3}x\nabla_\perp\boldsymbol{\cdot}\frac{n_{e}u_{e \parallel}m_{e}c^{2}}{2cB^{2}(\textbf{x})}\nabla_{\perp}\phi_{1}\hat{\chi}(\textbf{x})dt,
\end{equation}
considering that $ \int d\Omega_{gy} e F_{e}(x,v_{gy, \parallel})v_{gy, \parallel}=n_{e}u_{e \parallel}$. Assuming that $n_e u_{e \parallel}$ is constant in space, this results in

\begin{equation*}
\varepsilon_{\delta}^{2}\int\int d^{3}x\frac{u_{e \parallel}}{c}\frac{1}{8\pi}\frac{\omega_{pe}^{2}}{\Omega_{ce}^{2}}\nabla_{\perp}^{2}\phi_{1}\hat{\chi}(\textbf{x})dt= \varepsilon_{\delta}^{2}\int\int d^{3}x\frac{u_{e \parallel}}{c}\frac{\rho_{e}^{2}}{\lambda_{D}^{2}}\nabla_{\perp}^{2}\phi_{1}\hat{\chi}(\textbf{x})dt.
\end{equation*}
The term depending on $A_{1\parallel}$ becomes

\begin{equation}
=\frac{\varepsilon_{\delta}^{2}ec}{\Omega_{e}}\int\int d^{3}x\nabla_\perp \boldsymbol{\cdot} F_{e}(\textbf{x})\frac{1}{B(\textbf{x})}\nabla_{\perp}\left[\frac{v_{gy, \parallel}^{2}}{c^{2}} {A}_{1\parallel}(\textbf{x})\right]\hat{\chi}(\textbf{x})dtd\Omega_{gy}.
\end{equation}

Considering the electron distribution function to be close to a Maxwellian, such as 

\begin{equation}
    F_{e}=n_{0}\left(\frac{m_{e}}{2\pi}\right)^{3/2}\left(T_{0e\parallel}T_{0e\perp}^{2}\right)^{-1/2}\exp\left(-\frac{m_{e}v_{gy, \parallel}^{2}}{2T_{0e\parallel}}-\frac{m_{e}v_{\perp}^{2}}{2T_{0e\perp}}\right),
\end{equation}

we have

\begin{multline*}
    I=\int dVv_{\parallel}^{2}F_{e}=2\pi\int dv_{\perp}v_{\perp}\int dv_{\parallel}v_{\parallel}^{2}F_{e}\\
    =2\pi n_{0}\left(\frac{m_{e}}{2T_{0e\parallel}}\right)^{3/2}\left(T_{0e\parallel}T_{0e\perp}^{2}\right)^{-1/2}\frac{v_{the\perp}^{2}}{2}\frac{\sqrt{\pi}}{2}v_{the\parallel}^{3},
\end{multline*}

where we assumed $dV=v_{\perp}dv_{\perp}dv_{\parallel}d\theta $, and assumed that

\begin{equation*}
    \int_{0}^{\infty}dv_{\perp}v_{\perp}\exp\left(-\frac{v_{\perp}^{2}}{v_{the\perp}^{2}}\right)=-\frac{v_{the\perp}^{2}}{2}\left.\exp\left(-\frac{v_{\perp}^{2}}{v_{the\perp}^{2}}\right)\right|_{0}^{\infty}=\frac{v_{the\perp}^{2}}{2},
\end{equation*}

and 

\begin{equation}
    \int_{-\infty}^{\infty}dv_{\parallel}v_{\parallel}^{2}\exp\left(-\frac{v_{\parallel}^{2}}{v_{the\parallel}^{2}}\right)=\frac{\sqrt{\pi}}{2}v_{the\parallel}^{3},
\end{equation}

where we considered that

\begin{equation}
    \int_{-\infty}^{\infty}dxx^{2n}e^{-bx^{2}}=2\sqrt{\pi}\frac{\left(2n\right)!}{n!}\left(\frac{1}{2\sqrt{b}}\right)^{2n+1}.
\end{equation}

As a result, we have that

\begin{equation*}
    I=\frac{n_{0}}{2}\left(\frac{m_{e}}{2T_{0e\parallel}}\right)^{3/2}\frac{T_{0e\parallel}}{T_{0e\perp}}v_{t,e,\perp}^{2}v_{t,e,\parallel}^{3}  =\frac{n_{0}}{2}\frac{T_{0e\parallel}}{T_{0e\perp}}v_{t,e,\perp}^{2} =\frac{n_{0}T_{0e\parallel}}{m_{e}}.
\end{equation*}

This results in 

\begin{equation}
    \frac{e}{\Omega_{e}Bc} I = \frac{e}{\Omega_{e}Bc}\frac{n_{0}T_{0e\parallel}}{m_{e}} = \frac{n_{0}T_{0e\parallel}}{B^{2}}=\frac{\beta_{e}}{8\pi}.
\end{equation}

The final piece missing now is the fully kinetic contribution. The variation of the fully kinetic action with respect to the parallel magnetic potential give us 

\begin{equation*}
\frac{\delta}{\delta A_{1\parallel}(\textbf{x})}{\cal S}_{fk,i}^{p}\circ\hat{\chi}_{1}(\textbf{x})=\frac{\partial}{\partial\nu}\left[\int\int F_{i}(x,v)\frac{q_{i}}{c}\varepsilon_{\delta}(\textbf{A}_{1}(\textbf{x})+\nu\hat{\chi}(\textbf{x}))\dot{x}dtd\Omega\right]
\end{equation*}

\begin{equation}
=\frac{q_{i}}{c}\varepsilon_{\delta}\int\int F_{i}(x,v)\hat{\chi}(\textbf{x}) v_\parallel dtd\Omega= \frac{1}{c}n_i u_{i \parallel},
\end{equation}
where $u_{i \parallel}$  is the parallel ion velocity. Finally, considering all terms from the Lagrangian derived in this section, we have the parallel Ampere equation 

\begin{equation*}
\frac{1}{4\pi}\nabla^{2}{A}_{1 \parallel}(\textbf{x})+\frac{u_{e}}{c}\frac{\rho_{e}^{2}}{\lambda_{D}^{2}}\nabla_{\perp}^{2}\phi_{1}(\textbf{x})- \frac{\beta_{e}}{8\pi} \nabla_{\perp}^{2}{A}_{1\parallel}(\textbf{x})=
\end{equation*}

\begin{multline}
\frac{\varepsilon_{\delta}^{2}e^{2}}{mc^{2}}\int dW\bigl\langle F_{e}(\mathbf{x}-\rho)\bigr\rangle\textbf{A}_{1}(\textbf{x})  \\
-\varepsilon_{\delta}e \frac{1}{c}\int d\theta d\mu_{gy} v_{gy, \parallel} \left\langle F_{e}(\mathbf{x}-\mathbf{\mathbf{\rho}})\right\rangle \\
 +\sum_{i}\frac{q_{i}}{c}\int F_{i}(x,v)u_{i \parallel}d\Omega
\end{multline}
Using a low wave number approximation for the gyroaveraged electron distribution,  and taking in consideration the complete electromagnetic potential we have

\begin{equation}
\frac{c}{4\pi}\nabla_{\perp}^{2}{A}_{1\parallel}\left(1-\frac{\beta_{e}}{2}\right) - u_e\frac{\rho_{e}^{2}}{\lambda_{D}^{2}}\nabla_{\perp}^{2}\phi_{1}(\textbf{x})=\frac{e^{2}}{m_e c} n_{e}{A}_{1 \parallel}(\textbf{x})- I_{e}+\sum_ i I_{i \parallel},
\end{equation}

where $n_{i}=\int d\Omega F_{i}(x,v)$, and  $  \sum_{i}q_{i}\int F_{i}(x,v)u_{i \parallel}d\Omega=q_i\sum_{i}n_{i}u_{i \parallel} = \sum_{i } I_{i \parallel} $. We also considered $n_{e}=\int d\Omega F_{e}(\mathbf{x})$, and $I_{e}=e u_{e}n_e$. The first term on the right hand side is due to the velocity shift transformation performed in the gyrokinetic coordinates transformation detailed on the Appendix \ref{appendix_a}. This term is also associated with the magnetization current, which in this case comes out explicitly to account for the fact that for the electrons, the coordination reduction does not allow for the natural presence of diamagnetic current (\cite{Frank-Kamenetskii1972}).

\section{Linear Analysis}

Despite the vast complexity and myriad of physical phenomena taking place in the solar wind, in the plasma physics framework, an interesting and effective exercise to validate a model consists of performing a wave analysis. Waves are connected with energy transfer, but not necessarily dissipation, and are often present in turbulent systems (\cite{PhysRevResearch.2.043253}, \cite{Plasma_Astro}). In the present section we are going to perform a linear analysis in order to study the waves present in our model.

For that aim, we have modified the solver FIDEL (\cite{Felix}) and we will make a comparison with two other linear solvers, the fully kinetic one DSHARK (\cite{DSHARK}) and the hybrid kinetic-ion/fluid-electron code HYDROS (\cite{HYDROS}). We take a look at three main wave solutions that are of interest in the field of space physics, i.e. ion Bernstein waves, the limiting case of the highest frequency electrostatic lower hybrid wave and ion acoustic waves.

\subsection{Ion Acoustic Waves}

Ion acoustic waves (IAW) are a longitudinal oscillation of charged particles in the plasma similar to acoustic waves on neutral gas.  IAW are subject to strong Landau damping, that means that they represent an indispensable class of waves that could be involved in the energy dissipation mechanism of solar wind. They can also interact with the electromagnetic field. Such waves can be damped due to Coulomb collisions or Landau damping.

\begin{figure}
\begin{center}
\includegraphics[width=7.5cm]{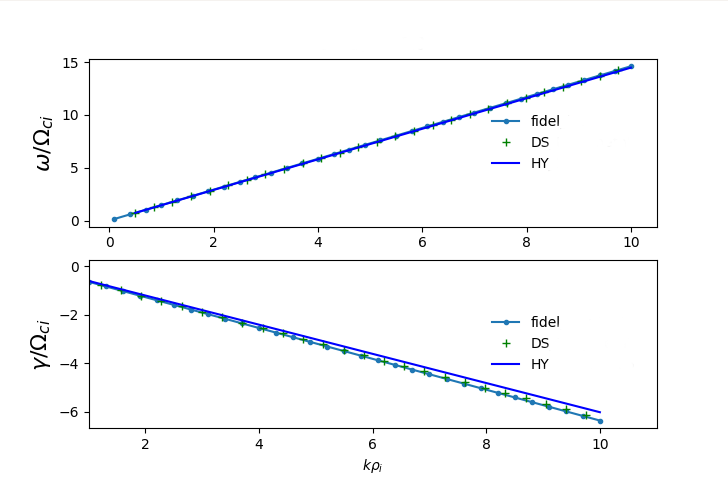}
\end{center}
\caption{Parallel ($\theta=5$) Ion Acoustic Wave frequency (top) and damping rate (bottom) for HYDROS (dark blue), DSHARK (green), and Fidel (light blue).  }\label{IAW_t5}
\end{figure}

In the present work, we took a look on IAW with a recently modified version of FIDEL, where we have included also the Maxwellian Laplacian in the Poisson equation, derived in the previous section. The Maxwellian Laplacian refers to the Laplacian on the Poisson equation that does not have a coefficient. This modification allows us to better study high frequency waves, a point that will be discussed in more detail. The linear property of the wave is observed in Figure \ref{IAW_t5}. For high values of wave number $k$, a difference between kinetic and fluid models is already observed. In this case, because the fluid framework does not capture the complete kinetic physics, perpendicular damping rates are not fully captured for low wavelength values. 

For a plasma with ion temperature smaller than electron temperature, and assuming a large wavelength limit, the dispersion relation takes the form of

\begin{equation}
\left ( \frac{\omega}{k} \right ) ^2 =\frac{ c_{s}^{2}}{1+k^{2} \lambda_{D e}^{2}}+3 \frac{\kappa_B T_{s 0}}{m_{s}},\label{IAW_analytic}
\end{equation}

where $c_s$, $\lambda_{D e}$, and $\kappa_B $, represents the sound speed, Debye length and Boltzmann constant, respectively. Details on the derivation for our model can be found in (\cite{Nathan}). It is important to emphasize that IAW propagates with constant velocity, where the velocity lies between the thermal velocities of the species present on the medium. Our model also allow us to take a look at parallel propagating IAW.

\begin{figure}
\begin{center}
\includegraphics[width=7.5cm]{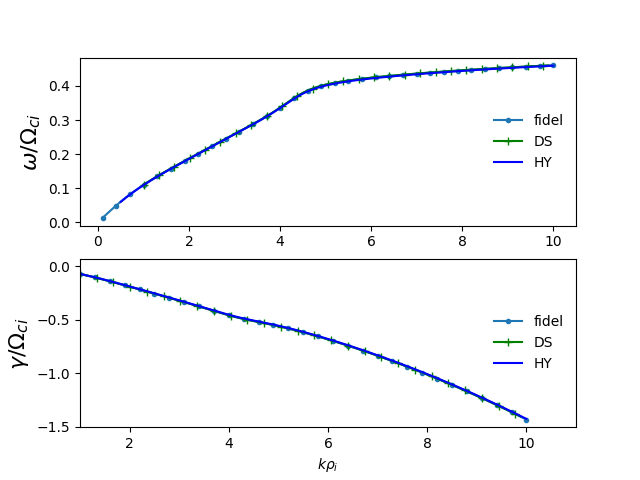}
\end{center}
\caption{Perpendicular ($\theta=85$) Ion Acoustic Wave frequency (top) and damping rate (bottom) for HYDROS (dark blue), DSHARK (green), and Fidel (light blue).  }\label{IAW_t85}
\end{figure}

The Landau damping of  IAW and  has been extensively discussed in (\cite{IAW_1}), where  Landau damping  strength is proportional to the ion thermal speed. The presence of IAW has also been associated with the increase of electron distribution anisotropy (\cite{IAW_3}), and also solar wind heating (\cite{IAW_4}).

\begin{figure}
\begin{center}
\includegraphics[width=6.5cm]{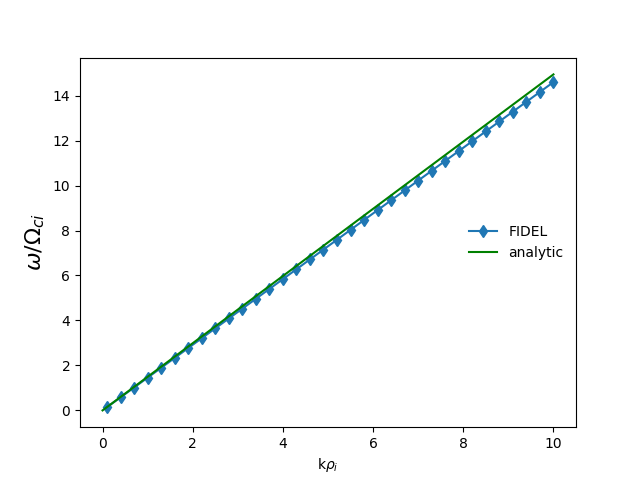}
\end{center}
\caption{Parallel ($\theta = 5$) comparison between the analytic IAW and FIDEL}\label{IAW_analytic_fig}
\end{figure}

In the limit of large wave lengths, that is  $k^{2} \lambda_{D e}^{2} \ll 1$, the dispersion relation depends only on the sound speed and the temperature ratio, i.e.  

\begin{equation}
\omega^2 = c_{s}^2 k^2  \left ( 1 +3 \frac{T_i}{T_e} \right ).
\end{equation}
With that in mind, we also take a look on how our model compares to the analytic solution, cf. Figure \ref{IAW_analytic_fig}. The agreement for low values of wave number is in accord with past FIDEL results. With the addition of the Maxwellian Laplacian, we are able to extend this agreement to higher values if $k$. As seen in Figure \ref{IAW_analytic_fig}, this agreement does not hold for higher values of $k$. This happens likely due to the limitation imposed by the reduced model. 

The presence of IAW modes in our model represent an important milestone in developing a reduced framework that is capable of, at a reduced computational cost, describe important kinetic phenomena which might be of great relevance to better understand the problem of energy dissipation in solar wind. 

\subsection{Ion Bernstein Waves} 

\begin{figure}
\begin{center}
\includegraphics[width=7.5cm]{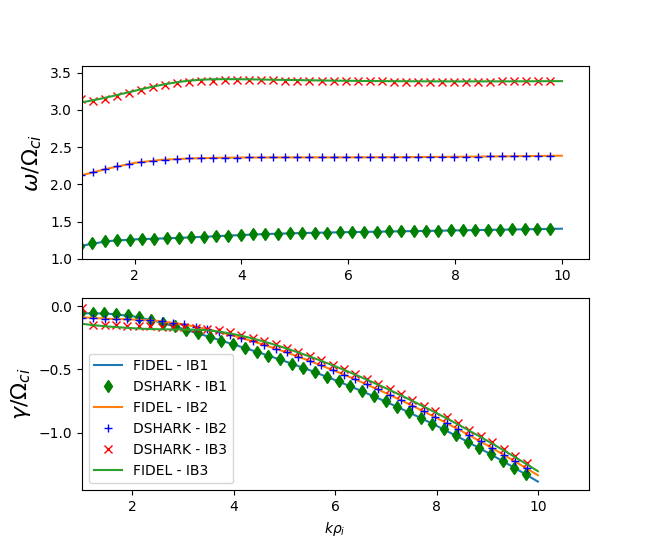}
\end{center}
\caption{Perpendicular ($\theta=85$) propagating Ion Bernstein Wave frequency (top) and damping rate (bottom) for different IBW modes}\label{IBW}
\end{figure}

Ion Bernstein waves (IBW) are slowly propagating, longitudinal hot plasma waves. Such waves interact with kinetic Alfven waves (KAW) (\cite{IBW_2}), and are subject to strong localized electron Landau damping (\cite{IBW_1}). Although their nature spurs out of the electromagnetic theory, it is interesting to observe that IBWs can already be observed in the present reduced model. In Figure \ref{IBW} we observe a comparison between FIDEL and DSHARK simulations. HYDROS simulations exhibit more unstable root-finder solutions and were therefore excluded from this particular analysis.    

The general format of our analytic IBW looks like 

$$
k_{\perp}^{2}+2\left(\frac{\omega_{p}}{v_{t}}\right)^{2}e^{-\frac{1}{2}k_{\perp}^{2}\rho_{i}^{2}}\sum_{n=-\infty}^{\infty}I_{n}(\frac{1}{2}k_{\perp}^{2}\rho_{i}^{2})\left[1+\zeta_{0}Z\left(\zeta_{n}\right)\right]=0,
$$
where $I_n$ stands for modified Bessel function and Z stands for the plasma function. More details about the derivation can be found in (\cite{Nathan}). 

\begin{figure}
\begin{center}
\includegraphics[width=7.5cm]{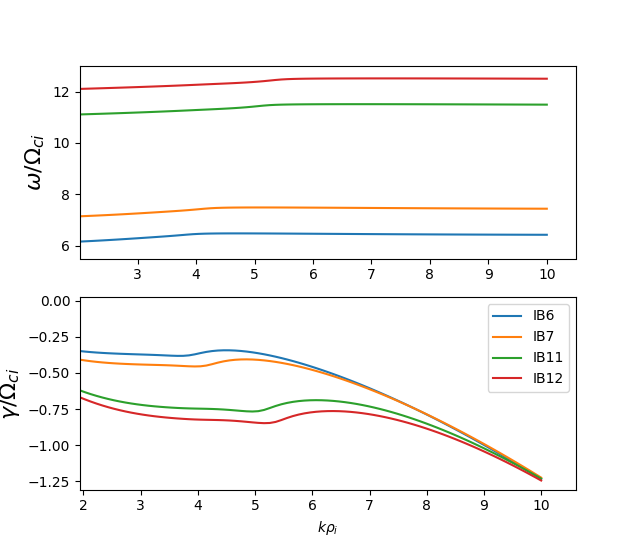}
\end{center}
\caption{Perpendicular ($\theta=85$) propagating Ion Bernstein Wave frequency (top) and damping rate (bottom) for higher IBW modes.  }\label{IBW_6}
\end{figure}

The agreement between both codes indicates that the reduced model is also capable of reproducing important fully kinetic linear features. It's imperative to have a model that can properly reproduce linear modes, as mode coupling between different wave solutions is believed to play a role in turbulence dissipation in collisionless plasmas.. Furthermore, the presently modified FIDEL solver is also capable of reproducing higher IBW modes, as seen in Figure \ref{IBW_6}, which is a  relevant point, also raised in (\cite{IBW_2}).

The presence of higher IBW modes is likely due to the presence of the Maxwellian Laplacian in the Poisson equation. As it will be discussed in more detail in the next section, the use of the Maxwellian Laplacian facilitates the computation of higher frequency waves in the system, which is of importance in the study of ion frequency range energy dissipation in solar wind turbulence. 

\subsection{Waves with $\omega \gg \Omega_{ci}$}

Turbulent heating is thought to be one of the sources of ion heating in solar wind evolution (\cite{high_freq_1}, \cite{high_freq_2}, \cite{high_freq_3}). Magnetic reconnection, Landau damping and transit time damping are thought to be the main processes involved in the dissipation mechanism at kinetic scales (\cite{high_freq_4}, \cite{high_freq_5}). The use of gyrokinetics to study solar wind turbulence has been  put into question due to the $\omega \ll \Omega_i$ restriction described in (\cite{IBW_2}). In the present model, we demonstrate that the drift kinetic limit of our hybrid kinetic-gyrokinetic formulation contains lower hybrid wave, demonstrating an important step in the direction of simulating KAW-like physics. 

When looking at high frequency waves, our model's dispersion relation takes the form 

\begin{equation}
\omega^{2}=\frac{1}{2}\left(\Omega_{ci}^{2}+\frac{\omega_{pe}^{2}k_{\parallel}^{2}+\omega_{pi}^{2}k^{2}}{c_{v}k^{2}+\frac{\omega_{pe}^{2}}{\Omega_{ce}^{2}{}_{\perp}}k_{\perp}^{2}}\right) \label{high_freq_eq}
\end{equation}
\[
+\frac{1}{2}\sqrt{\left(\Omega_{ci}^{2}+\frac{\omega_{pe}^{2}k_{\parallel}^{2}+\omega_{pi}^{2}k^{2}}{c_{v}k^{2}+\frac{\omega_{pe}^{2}}{\Omega_{ce}^{2}{}_{}}k_{\perp}^{2}}\right)^{2}+\frac{4(\omega_{pe}^{2}+\omega_{pi}^{2})\Omega_{ci}^{2}}{c_{v}k^{2}+\frac{\omega_{pe}^{2}}{\Omega_{ce}^{2}{}_{}}k_{\perp}^{2}}k_{\parallel}^{2}}.
\]

In this context, solutions  that satisfy $\omega \gg\Omega_{i}$ are compared with their analytical counterpart, in order to  examine whether or not the present model is capable of describing high frequency phenomena. In order to understand high frequency waves at various direction to the background magnetic field, we take a look on a $\theta$ scan. \\

\begin{figure}
\begin{center}
\includegraphics[width=8.5cm]{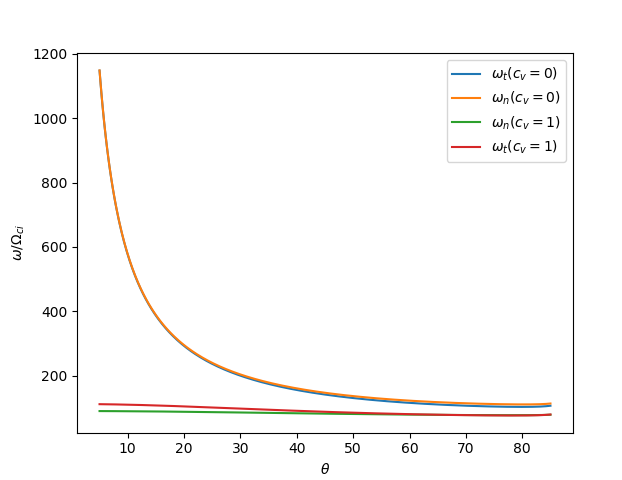}
\end{center}
\caption{$\theta$ scan at $k\rho_i=10$ and with $\frac{T_i}{T_e}=1$ of $\omega \gg \Omega_{ci}$.}\label{high_freq_t}
\end{figure}

Figure \ref{high_freq_t} shows us the high frequency wave \ref{high_freq_eq} with ($c_v=1$) and without ($c_v=0$) the Maxwellian Laplacian,  at $k\rho_i=10$ for various angles of propagation. It is clear at this point that at low $\theta$ values, a blow up of the frequency mode in the case without the Maxwellian Laplacian happens due to the fact that reduced models such as driftkinetic is valid to modes mainly perpendicular to the background magnetic field. The effect of the Maxwellian Laplacian is seen on the numerical ($\omega_n$, green) and theoretical ($\omega_t$, red) curves in Figure \ref{high_freq_t}. Nevertheless, at high frequencies we observe that the effect of the Maxwellian Laplacian supersede the effects of the electron polarization response. For parallel propagating waves, the frequency approaches the electron plasma frequency on the form of Langmuir oscillation. Assuming that solar wind density is around 3 $ cm^{-3}$, and the background magnetic field is around $10^{-5}$ Gauss (\cite{Verscharen}), we arrive at $\frac{\omega_{pe}}{\Omega_{ci}} \sim 90$. The main goal of the present model is to describe waves of the order of Langmuir waves in the $k \sim k_\parallel$ limit. For small values of $\theta$, both the numerical and theoretical linear limits of the model seen in Figure \ref{high_freq_t} predict values of the order of magnitude of the value computed using in situ data.

\section{Summary and Outlook}

Energy dissipation in the solar wind  has been regarded as one of the outstanding unsolved problems in space plasma physics (\cite{Goldstein2001}). While the full understanding of the various mechanisms that take place on the solar corona and solar wind are far from conclusive, it is expected that turbulence plays a major role (\cite{9_outstanding_problems}). The multi-scale nature of turbulence in space and astrophysics plasma imposes a considerable barrier in computing those systems from the injection to the dissipation range scale. In order to bridge this gap, reduced models are often used, and despite their constraints, a good agreement has been shown (\cite{Howes2006}, \cite{Told}). Despite that, in order to better understand energy dissipation at ion scale, one needs to take into account constraints that violate the gyrokinetic conditions (\cite{IBW_2}).  

With that in mind, we have derived a hybrid model that takes in consideration a reduced description for electrons, and ions are considered in their fully kinetic nature, in order to better compute ion scale range phenomena. We have seen how the reduced dynamics can be coherently coupled with the fully kinetic description through a coordinate transformation on the field equations frame of reference. Linearly, we have benchmarked our system with two dispersion solvers, HYDROS and DSHARK. We have seen that our system reproduces well not only ion acoustic waves and ion Bernstein waves, but also higher frequency waves. A discussion about the relationship between ion Bernstein waves and high frequency waves points to the direction that a reduced hybrid drift-kinetic electrons and fully kinetic ions is already capable of depicting higher frequency phenomena and therefore could be considered a step towards a reduced model capable of describing energy dissipation in solar wind. 

Nevertheless, the next steps are fundamental to cement the present results. The next stage consists of the implementation of an electromagnetic dispersion relation solver. That will allow us to study kinetic Alfven waves and their coupling with ion Bernstein waves, and when cross analysed with Landau damping (\cite{IBW_2}, \cite{PhysRevResearch.2.043253}) and ion Cyclotron damping (\cite{Chandran2010}) could shed some light in the understanding of one of the many dissipation mechanisms taking place in solar wind turbulence. Moreover, the nonlinear implementation of the system is already ongoing and could further provide support to the validity of the model. 

\section{Acknowledgments}

The authors thank Natalia Tronko for helpful support in the gyrokinetic derivation, Greg Hammet for insights and valuable discussion about the linear properties of the system, and Felix Gaisbauer for support in the transition from the old version of FIDEL to the current one.

\appendix

\section{}\label{appendix_a}

In the present appendix, instead of going over the whole gyrokinetic derivation performed for the present model, we are only going to point out the modifications we have implemented on the derivation performed in the literature, namely  (\cite{Littlejohn, littlejohn_1983}, \cite{Tronko2018}, \cite{Sugama}).

The standard gyrokinetic coordinate transformation consists of a series of mathematical maneuvers that aim at reducing the functional dependency of the Lagrangian of the system on the gyration angle. That functional dependency is expressed through the elimination of the $\theta$ variable through perturbation theory. 

The first part of the transformation consists of the perturbation generated due to the symmetry break imposed by a background magnetic field. That symmetry transform the fully kinetic Lagrangian 

\begin{equation*}
\Gamma=\left(m\textbf{v}+\frac{e}{c}\mathbf{A}(\textbf{x})\right)d\textbf{x} -Hdt,
\end{equation*}

where 

\begin{equation*}
H = \frac{m \textbf{v}^2}{2},
\end{equation*}

and $\mathbf{A}(\textbf{x})$, $\textbf{x}$, and $\textbf{v}$ stands for the magnetic potential, and the position and velocity coordinates, respectively. Since the strength of the symmetry break imposed by the background magnetic field is of a higher ordering, our model does not differ much from what is found on the literature at this point. This means that our Lagrangian remains with the same format (\cite{Nathan}) after the guiding center transformation, that is 

\begin{equation}
\mathcal{L}_{gc} =\left[\frac{e}{c}A(X_{gc})+mV_{\parallel}\hat{b}(X_{gc})-\frac{mc}{e}\mu R\right]\boldsymbol{\cdot} dX_{gc}+\frac{mc}{e}\mu d\Theta-H_{gc}dt\label{full_lagrangian_1or2},
\end{equation}
where

\begin{equation}
H_{gc} =\frac{1}{2}mV_\parallel^{2}-\mu B.\label{some_H}
\end{equation}

Our symplectic tensor becomes then

\begin{equation}
\omega_{gc}=d\Gamma_{gc}=\frac{e}{c}\epsilon_{ijk}B^{k}dX_{gc}^{i}\wedge dX_{gc}^{j}+m\hat{b}(X_{gc})dV_\parallel\wedge dX_{gc}^{i}+\frac{mc}{e}d\mu_{}\wedge d\theta_{},
\end{equation}
and in matrix form, it looks like

\begin{equation*}
\omega_{gc} = \left(\begin{array}{ccccccc}
 & dX_{gc}^{1} & dX_{gc}^{2} & dX_{gc}^{3} & dV_\parallel & d\mu & d\theta\\
dX_{gc}^{1} & 0 & \frac{e}{c}B_{3} & -\frac{e}{c}B_{2} & - mb_{1} & 0 & 0\\
dX_{gc}^{2} & -\frac{e}{c}B_{3} & 0 & \frac{e}{c}B_{1} & - mb_{2} & 0 & 0\\
dX_{gc}^{3} & \frac{e}{c}B_{2} & -\frac{e}{c}B_{1} & 0 & - mb_{3} & 0 & 0\\
dV_\parallel & mb_{1} & mb_{2} & mb_{3} & 0 & 0 & 0\\
d\mu & 0 & 0 & 0 & 0 & 0 & m\frac{c}{e}\\
d\theta & 0 & 0 & 0 & 0 & -m\frac{c}{e} & 0
\end{array}\right),
\end{equation*}
where the first row and column only indicate the corresponding coordinate from the wedge product. In order to transform the Hamiltonian $H_{gc} = \frac{1}{2}mV_\parallel ^{2}-\mu B(\textbf{X}_{gc})$, we construct the following vector field

\begin{equation}
i_{X_{H_{gc}}}\omega_{gc}+dH_{gc}=0.
\end{equation}
Similarly to what we have seen in the fully kinetic derivation, the transformation of the symplectic form induced a change of the Poisson structure, and considering any two functions $f,g\in(M_{gc},\mathbb{R})$ we have

\begin{equation}
\omega_{gc}(X_{f},X_{g})=\sum_{ab}\Pi_{gc}^{ab}(\Omega_{gc})\frac{\partial f}{\partial\Omega_{gc}^{a}}\frac{\partial g}{\partial\Omega_{gc}^{b}}.
\end{equation}

Since the relationship $\Pi_{gc}=\omega_{gc}^{-1}$ also holds on our gyrokinetic manifold, we have our new Poisson bracket

\begin{multline*}
\left\{ f,g\right\} _{gc}=\frac{e}{mc}\frac{1}{\epsilon_B}\left(\frac{\partial f}{\partial\theta}\frac{\partial g}{\partial\mu}-\frac{\partial f}{\partial\theta}\frac{\partial g}{\partial\mu}\right) \\
+\frac{B^{*}}{mB_{\parallel}^{*}}\left(\nabla^{*}f\frac{\partial g}{\partial V_{\parallel}}-\nabla^{*}g\frac{\partial f}{\partial V_{\parallel}}\right)-\epsilon_B\frac{c\hat{b}}{eB_{\parallel}^{*}}\left(\nabla^{*}f\times\nabla^{*}g\right) \label{gk_Poisson_bracket},
\end{multline*}

where the modified gradient is given by $\nabla^{*}=\nabla-\boldsymbol{R}^{*}\partial_{\theta_{}}$, $\boldsymbol{B}^{*}=\boldsymbol{B}+\frac{m}{e}cv_{\parallel}\nabla\times\hat{\boldsymbol{b}}-\frac{mc^{2}}{e^{2}}\mu\nabla\times \boldsymbol{R}^{*}$.  In order to compute the final guiding center coordinate transformation, we need to find the generator $S_{gc}$, which can be constructed in such a way that it eliminates the fluctuations on   $\tilde{H} = H - \left \langle H \right \rangle$. The Hamiltonian transformation is given by the Lie transformation

\begin{equation*}
H_{gc}(Z_{gc})=e^{-\mathcal{L}_{Sgc}}H(Z),
\end{equation*}

where 
\begin{equation}
\mathcal{L}_{S_{gc}} = \{ S_{gc}, \boldsymbol{\cdot} \}_{gc}.\label{Lie}
\end{equation}

Taking a look on the format of our Hamiltonian, and together with equation \ref{Lie}, the relationship for the generator should satisfy the cohomological equation
\begin{equation}\label{cohomological_equation}
\frac{eB}{mc}\frac{\partial S_{gc}}{\partial\theta}=(mV_{\parallel}\tilde{\mathcal{V}}_{\parallel}+mV_{\perp}\tilde{\mathcal{V}}_{\perp}),
\end{equation}

which give us 

\begin{equation*}
S_{gc}=\frac{m^{3}c^{2}}{e^{2}B^{2}}\left[\frac{V_{\parallel}V_{\perp}^{2}}{8}\left(\left(\hat{a}\boldsymbol{\cdot}\nabla\right)\hat{b}\boldsymbol{\cdot}\hat{a}-(\hat{c}\boldsymbol{\cdot}\nabla)\hat{b}\boldsymbol{\cdot}\hat{c}\right)+V_{\parallel}^{2}V_{\perp}(\nabla\times\hat{b})\boldsymbol{\cdot}\hat{a}+\frac{V_{\perp}^{3}}{3B}\hat{c}\boldsymbol{\cdot}\nabla B\right].
\end{equation*}

It is important to stress here that, in order to solve the cohomological equation \ref{cohomological_equation}, one must first apply the fluctuating operation on the given coordinate, i.e $\tilde{Z}_{gc} =  Z_{gc} - \left \langle Z_{gc} \right \rangle$, in the Hamiltonian. The coordinate transformation performed  now only affects the structure of the Hamiltonian, which means that we do not expect structural changes on the Lagrangian. In general terms, the new coordinate system, together with the last Lie transformation becomes

\begin{equation*}
z_{gc}=Z_{gc}+\sum_{n=1}\frac{\epsilon_{B}^{n}}{n!}\mathcal{Z}_{n,gc}(Z_{gc})+\left\{ S_{gc},Z_{gc}\right\} _{gc},\label{gc_coo}
\end{equation*}
where our Poisson bracket is constructed from the newly transformed symplectic component of the Lagrangian, and

\begin{equation*}
z_{gc}=\left[\begin{array}{c}
\textbf{x}_{gc}\\
\mathbf{v}_{s,\parallel}\\
\mathbf{v}_{s,\perp}\\
\theta
\end{array}\right],Z_{gc}=\left[\begin{array}{c}
X_{gc}\\
V_{\parallel}\\
V_{\perp}\\
\Theta
\end{array}\right],\mathcal{Z}_{gc}=\left[\begin{array}{c}
\chi\\
\mathcal{V}_{\parallel}\\
\mathcal{V}_{\perp}\\
\Omega
\end{array}\right],\{S_{gc},Z_{gc}\}=\left[\begin{array}{c}
(\hat{b}\boldsymbol{\cdot}\chi)\hat{b}-\frac{\hat{b}}{m}\frac{\partial S_{gc}}{\partial V_{\parallel}}\\
0\\
\frac{eB}{m^{2}V_{\perp}c}\frac{\partial S_{gc}}{\partial\Theta}\\
\frac{eB}{m^{2}V_{\perp}c}\frac{\partial S_{gc}}{\partial V_{\perp}}
\end{array}\right],
\end{equation*}
up to the ordering of our derivation. The second part of the coordinate transformation consists on the addition of the symmetry break due to the presence of electromagnetic perturbation. That transformation will give us our final gyrocenter coordination transformation. 

The gyrokinetic transformation consists of a new ordering for the electric field, that is $(k_\perp \rho_{th}) \frac{e\phi_1}{T_i}= \varepsilon_\perp \varepsilon_\delta$. Furthermore, in order to prevent large fluctuations of the magnetic field, we also consider an ordering for the magnetic potential, that means $\frac{\nabla A_1}{A_1} \sim \varepsilon_\delta^2$. For the full gyrokinetic derivation, $\varepsilon_\perp \sim 1$, that condition can be relaxed, in the case of a drift kinetic approximation, as it will be discussed in subsequent chapters. After a velocity shift and a Lie transformation, a new generating function is required in order to move towards the new coordinate system. 

At this point we need to perform a few modifications to the standard derivations. More details of this modification can be found at (\cite{Nathan}). For the sake of brevity, in this appendix we will content ourselves with laying down and understanding the reason behind the modifications.  

In constructing the new Hamiltonian, three Poisson brackets are required to find the equation for the first order generating function $S_1$ of the gyrokinetic Lie transformation, those are 

\begin{equation}
\{S_{1},H_{1}\}_{f}=\frac{e}{mc}(\frac{\partial S_{1}}{\partial\theta}\frac{\partial H_{1}}{\partial\mu}-\frac{\partial H_{1}}{\partial\theta}\frac{\partial S_{1}}{\partial\mu}),
\end{equation}

\begin{equation}
\{S_{1},H_{1}\}_{m}=\frac{B^{*}}{mB_{\parallel}}(\nabla^{*}S_{1}\frac{\partial H_{1}}{\partial v_{\parallel}}-\nabla^{*}S_{1}\frac{\partial S_{1}}{\partial v_{\parallel}})=\mathcal{O}(\varepsilon_{\delta}^{3}),
\end{equation}
and
\begin{equation}
\{S_{1},H_{1}\}_{s}=\frac{c\hat{b}}{eB_{\parallel}^{*}}(\nabla^{*}S_{1}\times\nabla^{*}H_{1}).
\end{equation}Here, $H_1$ stands for the perturbed Hamiltonian. In our model, due to the difference in strength of the fields acting on the gyrokinetic species, vis-à-vis the fully kinetic species, we can consider the last term equals to zero. This small change propagates itself in the derivation and a slightly modified final Lagrangian ensues. It looks like

\[
\Gamma_{gy}= \left(\frac{e}{\varepsilon_\delta c}\textbf{A}_{1}+m\textbf{v}_{gy, \parallel }\hat{\textbf{b}}(X_{gy})\right)\boldsymbol{\cdot}\dot{\textbf{X}}_{gy}+\varepsilon_\delta\frac{mc}{e}\mu_{gy}\dot{\theta}_{gy}-
\]
\begin{equation*}
\frac{1}{2}mv_{gy, \parallel}^{2}-\mu_{gy}B(\textbf{X}_{gy})-\varepsilon_{\delta}e\langle\psi_{1}\rangle-\varepsilon_{\delta}^{2}e^{2}\left(\frac{1}{2mc^{2}}\langle|\textbf{A}_{1}|^{2}\rangle-\frac{1}{2B(\textbf{X}_{gy})}\partial_{\mu_{gy}}\langle\psi_{1}^{2}\rangle\right),
\end{equation*}
where our new coordinate system is described by

\begin{equation}
Z_{gy}=z_{gc}+\varepsilon_{\delta}\left\{ S_{gy},z_{gc}\right\} _{gc}\label{gy_coo},
\end{equation}

where, up to the ordering of our derivation, we have that

\begin{equation*}
Z_{gy}=\left[\begin{array}{c}
X_{gy}\\
v_{gy,\parallel}\\
\mu_{gy}\\
\theta_{gy}
\end{array}\right],z_{gc}=\left[\begin{array}{c}
X_{gc}\\
v_{s,\parallel}\\
\mu\\
\theta
\end{array}\right],\{S_{gy},Z_{gc}\}=\left[\begin{array}{c}
-\left(\frac{c\hat{b}}{eB_{\parallel}^{*}}\times\nabla S_{1}+\frac{B^{*}}{mB_{\parallel}^{*}}\frac{\partial}{\partial v_{s,\parallel}}\int^{\theta}\tilde{H}_{1}d\theta\right)\\
0\\
\frac{e}{B}\tilde{H}_{1}\\
-\frac{1}{B}\frac{\partial}{\partial\mu}\int^{\theta}\tilde{H}_{1}d\theta
\end{array}\right].
\end{equation*}

To summarize the modifications discussed in the present appendix, we would like to point out that, whilst some of the old literature (\cite{Littlejohn}) only performs a guiding center approach, we follow more recent approaches and perform a further transformation due to the electromagnetic perturbation fields, i.e. the gyrokinetic transformation. Furthermore, similarly to \cite{Tronko2018}, we make use of an ordering associated to the curvature of the background magnetic field, as opposed to LittleJohn's, which assumes an arbitrary $\epsilon$, where the asymptotic derivation is performed by taking $\epsilon \rightarrow 0$. Our backgrounds magnetic field is considered to be slab, which further differs our model from \cite{Tronko2018}.

Specifically, our model deviates from the approach taken in previous works by further applying the Darwin approximation (\cite{Darwin}) in the calculation of the magnetic field's contribution, and not only on the electric field. In the Darwin approximation, the time derivative of the fields on a Coulomb gauge Lagrangian is neglected. In the context of the present work, that implies that there are no displacement current on the system. Furthermore, our work differs from \cite{Tronko2018}, \cite{Tronko}, and \cite{Tronko_2019} on the use of a slab geometry approximation for the background magnetic field, which is more suitable for space and astrophysical plasma, as well as minor modifications on the second coordinate transformation, as discussed above. 

Moreover, our model further derive the strong version of the field equations. Such a derivation is more suitable for a grid based numerical implementation. While we draw the inspiration to the field theoretical approach from works such as \cite{Sugama}, \cite{Brizard}, and \cite{littlejohn_1983}, our paths diverge at the computation of the final equations, where instead of adhering to the standard approach, we make use of a coordinate transformation embedded in a variational derivative. More details on this derivation are also found on the main body of the present manuscript. \\

\section{}\label{appendix_b}

We are going to perform the following derivative
\begin{equation*}
\partial_{\mu_{gy}}\left\langle \tilde{\psi}_{1}^{2}\right\rangle =\partial_{\mu_{gy}}\frac{1}{2\pi}\int d\theta\left[\psi_{1}(\boldsymbol{X}_{gy}+\boldsymbol{\rho})-\left\langle \psi_{1}(\boldsymbol{X}_{gy}+\boldsymbol{\rho})\right\rangle \right]^{2}
\end{equation*}

\begin{equation}
=\partial_{\mu_{gy}}\frac{1}{2\pi}\int d\theta\left[\psi_{1}(\boldsymbol{X}_{gy}+\boldsymbol{\rho})-\frac{1}{2\pi}\int d\theta'\psi_{1}(\boldsymbol{X}_{gy}+\boldsymbol{\rho}+\boldsymbol{\rho}')\right]^{2}.
\end{equation}

We must first define 

\begin{equation*}
\boldsymbol{\rho}=\rho\hat{\boldsymbol{\rho}}=\rho(cos\theta\boldsymbol{e}_{1}-sin\theta\boldsymbol{e}_{2}),
\end{equation*}
and
\begin{equation}
\boldsymbol{\rho}'=\rho\hat{\boldsymbol{\rho}'}=\rho(cos\theta'\boldsymbol{e}_{1}-sin\theta'\boldsymbol{e}_{2}),
\end{equation}

where $\rho=\frac{1}{\Omega_{e}}\sqrt{\frac{2\mu_{gy}B(X_{gy})}{m_{e}}}$ and$ \partial_{\mu_{gy}}v_{\perp}=c/e\rho$, we can then write
\begin{equation*}
\frac{\partial\boldsymbol{\rho}}{\partial\mu_{gy}}=\frac{1}{\Omega_{e}}\frac{c}{e\rho}=\sqrt{\frac{m_{e}c^{2}}{2e\mu_{gy}B(X_{gy})}},
\end{equation*}
and
\begin{equation*}
\partial_{\mu_{gy}}\psi_{1}(\boldsymbol{X}_{gy}+\boldsymbol{\rho})=\nabla\psi_{1}\boldsymbol{\cdot}\frac{\partial\boldsymbol{\rho}}{\partial\mu_{gy}}=\nabla\psi_{1}\boldsymbol{\cdot}\hat{\boldsymbol{\rho}}\frac{1}{\Omega_{e}}\frac{c}{e\rho},
\end{equation*}
which we can also write as
\begin{equation}
\partial_{\mu_{gy}}\psi_{1}(\boldsymbol{X}_{gy}+\boldsymbol{\rho}+\boldsymbol{\rho}')=\nabla\psi_{1}\boldsymbol{\cdot}\frac{\partial(\boldsymbol{\rho}+\boldsymbol{\rho}')}{\partial\mu_{gy}}=\nabla\psi_{1}\boldsymbol{\cdot}(\boldsymbol{\rho}+\boldsymbol{\rho}')\frac{1}{\Omega_{e}}\frac{c}{e\rho}.
\end{equation}

Using the product rule on the quadratic term of the electromagnetic potential, we have

\begin{equation}
\partial_{\mu_{gy}}\left\langle \tilde{\psi}_{1}^{2}\right\rangle =\left\langle \partial_{\mu_{gy}}\tilde{\psi}_{1}^{2}\right\rangle =2\left\langle \tilde{\psi_{1}}\partial_{\mu_{gy}}\tilde{\psi}_{1}\right\rangle 
\end{equation}

\begin{equation}
=2\left\langle \tilde{\psi_{1}}\nabla\psi_{1}\boldsymbol{\cdot}\hat{\boldsymbol{\rho}}\frac{1}{\Omega_{e}}\frac{c}{e\rho}\right\rangle -2\left\langle \tilde{\psi_{1}}\left\langle \nabla\psi_{1}\boldsymbol{\cdot}(\boldsymbol{\rho}+\boldsymbol{\rho}')\frac{1}{\Omega_{e}}\frac{c}{e\rho}\right\rangle \right\rangle .
\end{equation}

Explicitly describing the non fluctuating electromagnetic potential, we have 

\begin{equation*}
\partial_{\mu_{gy}}\left\langle \tilde{\psi}_{1}^{2}\right\rangle =\left\langle \partial_{\mu_{gy}}(\psi_{1}-\left\langle \psi_{1}\right\rangle )^{2}\right\rangle =\left\langle \partial_{\mu_{gy}}(\psi_{1}^{2}-2\psi_{1}\left\langle \psi_{1}\right\rangle -\left\langle \psi_{1}\right\rangle ^{2})\right\rangle .
\end{equation*}

Expanding the electromagnetic potential give us 

\begin{equation*}
\psi_{1}(\textbf{X}_{gy}+\boldsymbol{\rho})=\psi_{1}(\textbf{X}_{gy})+\boldsymbol{\rho}\boldsymbol{\cdot}\nabla\psi_{1}(\textbf{X}_{gy}).
\end{equation*}

We observe that a few relationships can be drawn from this expansion and previous derivations, including 

\begin{equation}
\left\langle \psi_{1}(\textbf{X}_{gy}+\boldsymbol{\rho})\right\rangle =\psi_{1}(\textbf{X}_{gy})+\left\langle \boldsymbol{\rho}\right\rangle \boldsymbol{\cdot}\nabla\psi_{1}(\textbf{X}_{gy})=\psi_{1}(\textbf{X}_{gy}),
\end{equation}

\begin{equation*}
\left\langle \tilde{\psi}_{1}(\textbf{X}_{gy}+\boldsymbol{\rho})\right\rangle =0,
\end{equation*}
and
\begin{equation*}
\left\langle \psi_{1}^{2}(\textbf{X}_{gy}+\boldsymbol{\rho})\right\rangle =\left\langle \psi_{1}^{2}-2\psi_{1}\left\langle \psi_{1}\right\rangle -\left\langle \psi_{1}\right\rangle ^{2}\right\rangle.
\end{equation*}
We have then

\begin{equation*}
\left\langle \psi_{1}^{2}(\textbf{X}_{gy}+\boldsymbol{\rho})\right\rangle =\left\langle \psi_{1}(\textbf{X}_{gy})+\boldsymbol{\rho}\boldsymbol{\cdot}\nabla\psi_{1}(\textbf{X}_{gy})\right\rangle ^{2}
\end{equation*}

\begin{equation*}
=\psi_{1}^{2}(\textbf{X}_{gy})+\left\langle [\boldsymbol{\rho}\boldsymbol{\cdot}\nabla\psi_{1}(\textbf{X}_{gy})]^{2}\right\rangle 
\end{equation*}

\begin{equation*}
=\psi_{1}^{2}(\textbf{X}_{gy})+\frac{1}{2\pi}\int d\theta[\rho(cos\theta\boldsymbol{e}_{1}-sin\theta\boldsymbol{e}_{2})\boldsymbol{\cdot}\nabla\psi_{1}(\textbf{X}_{gy})]^{2}
\end{equation*}

\begin{equation*}
=\psi_{1}^{2}(\textbf{X}_{gy})+\frac{\rho^{2}}{2}((\partial_{1}\psi_{1})^{2}+(\partial_{2}\psi_{1})^{2})-\frac{\rho^{2}}{2\pi}\partial_{1}\psi_{1}\partial_{2}\psi_{1}\int d\theta cos\theta sin\theta
\end{equation*}

\begin{equation}
=\psi_{1}^{2}(\textbf{X}_{gy})+\frac{\rho^{2}}{2}\left|\nabla_{\perp}\psi_{1}(\textbf{X}_{gy})\right|^{2}.
\end{equation}
Furthermore 

\begin{equation}
\left\langle \psi_{1}\left\langle \psi_{1}\right\rangle \right\rangle =\frac{1}{2\pi}\int d\theta[\psi_{1}(\textbf{X}_{gy}+\boldsymbol{\rho})\psi_{1}(\textbf{X}_{gy}+\boldsymbol{\rho})]=\psi_{1}^{2}(\textbf{X}_{gy}).
\end{equation}

Where we assumed that $ \rho+\rho'\ll$ L. Finally, the averaged squared perturbed electromagnetic potential becomes

\begin{equation}
\left\langle \tilde{\psi}_{1}^{2}(\textbf{X}_{gy}+\boldsymbol{\rho})\right\rangle =\psi_{1}^{2}(\textbf{X}_{gy})+\frac{\rho^{2}}{2}\left|\nabla_{\perp}\psi_{1}^{2}\right|-2\psi_{1}^{2}(\textbf{X}_{gy})+\psi_{1}^{2}(\textbf{X}_{gy})=\frac{\rho^{2}}{2}\left|\nabla_{\perp}\psi_{1}\right|^{2},
\end{equation}

and considering that $\partial_{\mu_{gy}}\rho^{2}=2\rho\partial_{\mu_{gy}}\rho$, we have then

\begin{equation}
\partial_{\mu_{gy}}\left\langle \tilde{\psi}_{1}^{2}(\textbf{X}_{gy}+\boldsymbol{\rho})\right\rangle =\rho\frac{\partial\rho}{\partial\mu_{gy}}\left|\nabla_{\perp}\psi_{1}(\textbf{X}_{gy})\right|^{2}=\frac{c}{e\Omega_{e}}\left|\nabla_{\perp}\psi_{1}(\textbf{X}_{gy})\right|^{2}.
\end{equation}

\section{}\label{appendix_c}

In order to perform a preliminary numerical analysis of our system of equations, we are going to take a look at a simpler case. In this section, we are going to construct the Vlasov equation using the equations of motion derived in the section four of the present manuscript, and we will perform a drift kinetic and electrostatic approximation to the electron species of the system. The equation reads
 \begin{equation}
     \frac{\partial F}{\partial t} + \dot{X}_{gy} \boldsymbol{\cdot} \nabla_{gy} F + \dot{v}_{gy, \parallel} \partial_{\bar{v}_{gy, \parallel}} F = 0,
 \end{equation}
 
that is

\begin{equation*}
   \frac{\partial F}{\partial t} +  \frac{\textbf{B}^{*}}{mB_{\parallel}^{*}} \left \{ m {v}_{gy, \parallel} - \varepsilon_\delta \frac{e}{c} \left < A_{1 \parallel }\right > \right \} \boldsymbol{\cdot} \nabla_{gy} F
\end{equation*}
  
\begin{equation*}
     + \frac{c \hat{b}}{e B^*_{\parallel}} \times \left ( \mu_{gy} \nabla_{gy} B (X_{gy}) + \varepsilon_\delta e \nabla \left < \phi_1 \right > -\varepsilon_\delta \frac{e}{c}{v}_{gy, \parallel} \nabla \left < A_{1 \parallel }\right >  \right ) \boldsymbol{\cdot} \nabla_{gy} F
\end{equation*}
      
\begin{equation}\label{25}
      - \frac{\textbf{B}^{*}}{mB_{\parallel}^{*}} \boldsymbol{\cdot} \left ( \mu_{gy} \nabla_{gy} B(X_{gy}) + e \varepsilon_\delta \nabla \left < \phi_1 \right > - \frac{e}{c} \bar{v}_{gy, \parallel} \varepsilon_\delta \nabla \left < A_{1 \parallel} \right >  \right ) \partial_{{v}_{gy, \parallel}} F = 0 
\end{equation}
      
In the first line we have the term that represents the parallel motion. The first term in the second line represents the grad-B drift, the second one is the E$\times$B drift, and the last term in the same line is the B flutter. The last line is associated with the parallel acceleration.

First we consider that $A_1$ vanishes, the Vlasov equation becomes then
 
\begin{equation}
   \frac{\partial F}{\partial t} + \frac{\textbf{B}^{*}}{mB_{\parallel}^{*}} m {v}_{gy, \parallel} \boldsymbol{\cdot}  \nabla_{gy} F +  e  \frac{c \hat{b}}{e B^*_{\parallel}} \times \nabla \left < \phi_1 \right > \boldsymbol{\cdot}  \nabla_{gy} F -  \frac{\textbf{B}^{*}}{mB_{\parallel}^{*}} \boldsymbol{\cdot} \left ( e  \nabla \left < \phi_1 \right > \right ) \frac{ \partial F }{ \partial {v}_{gy, \parallel}} = 0.
\end{equation}

Considering that $\textbf{B}^* = \textbf{B} + \frac{mc}{e} \bar{v}_{gy, \parallel} \nabla \times \hat{b} - \frac{mc^2}{e^2} \mu_{gy} \nabla ^* R^*$, and unless otherwise explicit, considering also that

\begin{equation}
    \frac{\textbf{B}^*}{B^*_\parallel} =  \frac{\textbf{B}^*}{B + \frac{m}{e}c {v}_{gy, \parallel} (\hat{b} \boldsymbol{\cdot} \nabla \times \hat{b}) - \frac{mc^2}{e^2} \mu_{gy} (\hat{b} \boldsymbol{\cdot} \nabla^*R^*)} \approx \hat{b},
\end{equation}

we have

 \begin{equation*}
\frac{\partial F}{\partial t}+{v}_{gy, \parallel}\partial_{z,gy}F+e\frac{c\hat{b}}{eB^*_{\parallel}}\times \partial_z\left<\phi_{1}\right> \boldsymbol{\cdot} \partial_{z,gy}F-\left(\frac{e}{m}\partial_z\left<\phi_{1}\right>\right) \partial_{v_{gy, \parallel}} F,
  \end{equation*}
  
which means we can write our main Vlasov equations as 
  
 \begin{equation}
\frac{\partial F}{\partial t}+{v}_{gy, \parallel}\partial_{z,gy}F+\frac{c}{B_{\parallel}}\left ( \hat{b}\times \partial_z\left<\phi_{1}\right>\right ) \boldsymbol{\cdot} \partial_{z,gy}F-\left(\frac{e}{m}\partial_z\left<\phi_{1}\right>\right)\partial_{v_{gy, \parallel}} F=0.\label{drift_kinetic_vlasov}
  \end{equation}

Now we are going to perform a linearization of the Vlasov and field equations, and proceed with the study of the modified electrostatic dispersion solutions of the FIDEL code (\cite{Felix}).

\subsection{Driftkinetic electrons}

We proceed with the linearization of the drift kinetic Vlasov equation, which is responsible for the electron dynamics. For pedagogic sake, we are gonna start the linearization from the very first equation \ref{25}, we have then 

 \begin{equation}
\frac{\partial F_e}{\partial t}+{v}_{gy, \parallel}\boldsymbol{\cdot}\nabla_{gy}F_e+\frac{c}{B^*_{\parallel}}\left ({{{\hat{b}}}}\times \nabla\left<\phi_{1}\right>\right )\boldsymbol{\cdot}\nabla_{gy}F_e-\left(\frac{e}{m_e} {\hat{b}} \boldsymbol{\cdot} \nabla\left<\phi_{1}\right>\right) \partial_{v_{gy, \parallel}} F_e =0.
  \end{equation}
The linearization is performed by working independently with each one of the terms in equation \ref{drift_kinetic_vlasov}.  First we start with the decomposition of the distribution function such as
 \begin{equation*}
\frac{\partial F}{\partial t} = \frac{\partial}{\partial t} \left ( F_0 + \delta F \right ) = -i\omega \delta F,
\end{equation*}

where we considered $F = F_0 + \delta F e^{i(\textbf{k} \boldsymbol{\cdot} \textbf{x} - wt)}$ and $F_0$ to be a spatially homogeneous Maxwellian. Using the parallel component of $\nabla$ on the canonical moment, we have

 \begin{equation*}
\frac{\textbf{B}^*}{mB_\parallel^*} m {v}_{gy, \parallel} \nabla_{\parallel gy} \delta F_e =  \frac{\textbf{B}^*}{B_\parallel^*} {v}_{gy, \parallel} i k_\parallel \delta F_e,
\end{equation*}

and the second term becomes
 \begin{equation*}
\frac{e\textbf{B}^*}{mcB_\parallel^*} \left < A_{1 \parallel }\right > \boldsymbol{\cdot} \nabla_{\parallel gy} \delta F_e = \frac{\Omega_{ce}}{B_\parallel ^ *} \delta A_{1 \parallel} i k_\parallel \delta F_e .
\end{equation*}

The first term of the second line of \ref{25} is related to the $\nabla B$  drift, and becomes 

\begin{equation}
\frac{c \hat{b}}{e B_\parallel} \times \mu_{gy} \nabla_{gy} B(X_{gy}) \boldsymbol{\cdot}  \nabla_{gy}F_e = \frac{c \hat{b}}{e B_\parallel} \times \mu_{gy} \nabla_{gy} B(X_{gy}) \boldsymbol{\cdot} \nabla_{ \perp gy} \delta F_e = 0.
\end{equation}

The $E \times B$ term becomes 

\begin{equation}
\frac{c \hat{b}}{e B_\parallel} \times e \nabla \left < \phi_1 \right > \boldsymbol{\cdot} \nabla_{gy}F_e = \frac{c}{B_\parallel} \hat{b} \times \nabla_\perp \phi_1 \boldsymbol{\cdot} \nabla_\perp \delta F_e =  0,
\end{equation}

and the magnetic flutter develops into

\begin{equation}
\frac{c \hat{b}}{e B_\parallel} \times \frac{e}{c} {v}_{gy \parallel} \nabla \left < A_{1 \parallel} \right > \boldsymbol{\cdot} \nabla_{gy}F_e = \frac{\hat{b}}{B_\parallel} \times  {v}_{gy \parallel} \nabla_\perp A_{1 \parallel} \boldsymbol{\cdot} \nabla_\perp \delta F_e = 0.
\end{equation}

The last line on equation \ref{25} is related to the parallel acceleration. For the first term on the same line we have that

\begin{equation*}
\frac{1}{m} \mu_{gy} \frac{\partial}{\partial z} B(X_{gy}) \partial_{v_{gy,\parallel}}F_e = 0.
\end{equation*}

The second term is 
\begin{equation}
\frac{{\textbf{B}}}{mB^*_\parallel} \boldsymbol{\cdot} e \nabla \left < \phi_1 \right > \partial_{v_{gy,\parallel}}F_e =\frac{e{\textbf{B}}}{mB^*_{\parallel}}\boldsymbol{\cdot} i{k}\delta\phi\partial_{v_{gy,\parallel}}F_{0,e} = \frac{e}{m} i{k_\parallel}\delta\phi\partial_{v_{gy,\parallel}}F_{0,e},
\end{equation}

and the third term becomes zero in the electrostatic limit. It is important to remember also that, for solar wind turbulence study, a slab approximation suffices to model the geometry of the magnetic field, the final linearized equation becomes then

\begin{equation}
 -i\omega \delta F_e +  {v}_{gy, \parallel} i k_\parallel \delta F_e - \frac{e}{m} i{k}\delta\phi\partial_{v_{gy,\parallel}}F_{0,e} = 0.\label{electron_vlasov_linear}
\end{equation}
The distribution function is computed as solutions of the system, and its integral give us the electron perturbation density, $\delta n_e$, which will be in turn used in the linearized Poisson equation. We again refer the reader to  (\cite{HYDROS}, \cite{Nathan}, \cite{Felix}) for further details.

\subsection{Vlasov ions}

We start with the linearization of the ion Vlasov equation. From equation \ref{final_fully} we have that

\begin{equation}
\frac{\partial F}{\partial t}+ \textbf{v}  \boldsymbol{\cdot} \nabla F+\frac{e}{m_{}}\left(\nabla \phi +\textbf{v} \times \textbf{B} \right) \boldsymbol{\cdot} \nabla_{v}F=0\label{ion_vlasov_1st_apndx}.
\end{equation}

We linearise equation \ref{ion_vlasov_1st_apndx} term by term, and then proceed to perform a plane wave assumption, such as 

\begin{equation}
 F = F_0 + \delta F e^{i \left( \textbf{k} \boldsymbol{\cdot} \textbf{x} -\omega t\right)}
\end{equation}
The first term becomes then 

\begin{equation}
\frac{\partial F}{\partial t}=\frac{\partial}{\partial t}\left(F_{0}+\delta F\right)=-i\omega\delta F.
\end{equation}
The second and third terms become then

\begin{equation}
{\textbf{v}}\boldsymbol{\cdot}\nabla F=i{\textbf{v}\boldsymbol{\cdot}\boldsymbol{k}}\delta F,
\end{equation}
\begin{equation}
-\frac{e}{m}\nabla\phi\boldsymbol{\cdot}\nabla_{v}F=\frac{e}{m}i\boldsymbol{k}\delta\phi\boldsymbol{\cdot}\nabla_{v}F_{0}.
\end{equation}

Finally, the last term becomes

\begin{equation}
\textbf{v} \times \textbf{B} \boldsymbol{\cdot} \nabla_{v}\delta F = 0 .
\end{equation}

Considering all terms together, and considering a standard Maxwellian distribution for $F_0$, our linearized ion Vlasov equation takes the form

\begin{equation}
-i\omega\delta F + i{\textbf{v}\boldsymbol{\cdot}\boldsymbol{k}}\delta F -\frac{e}{m}i\boldsymbol{k}\delta\phi\boldsymbol{\cdot}\nabla_{v}F_{0}=0,\label{ion_vlasov_linear}
\end{equation}
the solutions for this system are integrated and we can then construct the linear ion density perturbation $\delta n_i$. The solution for this equations is rather complex and its derivation rather cumbersome, for detail on this, we refer the reader to (\cite{HYDROS}, \cite{Felix}), and more specifically to (\cite{Nathan}) .

\subsection{Poisson equation}
In order to perform the linearization of the Poisson equation, we are going to work with the following  considerations:

\begin{equation}
\frac{1}{4\pi}\nabla^{2}\phi_{1}\left(\textbf{x}\right)+\frac{\rho_{th}^{2}}{\lambda_{D}^{2}}\nabla_{\perp}^{2}\psi_{1}\left(\textbf{x}\right)=\sum_{i}q_{i}\eta_{i}\left(\textbf{x}\right)-e\eta_{e}\left(\textbf{x}\right),
\label{apndx_poisson}
\end{equation}

\begin{equation*}
\psi_{1}\left(\textbf{X}_{gy}+\boldsymbol{\rho}\right) \sim \phi_{1},
\end{equation*}
\begin{equation*}
\phi_{1}\left(x\right)=\tilde{\phi_{1}}\exp\left[i\left(\textbf{k}\boldsymbol{\cdot}\textbf{x}-\omega t\right)\right].
\end{equation*}

The left-hand side of equation \ref{apndx_poisson} becomes

\begin{equation}
\frac{1}{4\pi}\nabla^{2}\phi_{1}\left(x\right)+\frac{\rho_{th}^{2}}{\lambda_{D}^{2}}\nabla_{\perp}^{2}\psi_{1}\left(x\right)=
\end{equation}
\[
\frac{1}{4\pi}\left(-k^{2}\tilde{\phi_{1}}\exp\left[i\left(\textbf{k}\boldsymbol{\cdot}\textbf{x}-\omega t\right)\right]\right)+ \frac{\rho_{th}^{2}}{\lambda_{D}^{2}}\left(-k_{\perp}^{2}\tilde{\phi_{1}}\exp\left[i\left(\textbf{k}\boldsymbol{\cdot}\textbf{x}-\omega t\right)\right] \right)
\]

\[
=\left(-\frac{1}{4\pi}k^2-\frac{\rho_{th}^{2}}{\lambda_{D}^{2}}k_{\perp}^{2}\right)\Tilde{\phi}_{1}.
\label{eq102}
\]

Due to background quasi neutrality, the right hand side of equation (\ref{apndx_poisson}) can be expressed as

\begin{equation*}
\sum_{i}q_{i}\eta_{i}\left(x\right)-e\eta_{e}\left(x\right) = q_i \delta n_i + q_e \delta n_e,
\end{equation*}
and our  linearized Poisson equations becomes

\begin{equation}
- \left( k^2+\frac{1}{2}\frac{\omega_{pe}^2}{\Omega_{ce}^2}k_\perp^2\right)\phi_1=4\pi\left(q_i \delta n_i+q_e \delta n_e\right),
\label{poisson_linear}
\end{equation}
where $\delta n_i$ is the perturbed ion density and $ \delta n_{e}$ is the perturbed electron density computed in the guiding center position, also known as the gyrocenter density distribution. 

The equations \ref{ion_vlasov_linear}, \ref{electron_vlasov_linear}, and \ref{poisson_linear} form a group of linear equations that must be solved simultaneously. In order to find the slutions of the system, we solve the perturbed ion and electron equations, and by integrating them over velocity space, we then are able to sove the field equation \ref{poisson_linear}. Such a solution gives us the wave dispersion relation and it takes the form

\begin{equation}
D(k,\omega) \delta \phi = 0,\label{D}
\end{equation}

where $D(k,\omega)$ is in general the determinant of the matrix of the system. The solutions of equation \ref{D} provide us with the various waves we aim to study.

\bibliographystyle{jpp}

\bibliography{jpp-instructions}

\end{document}